\documentclass[runningheads]{llncs}

\usepackage{lineno}
\usepackage{xcolor}
\usepackage{graphicx}
\usepackage{enumitem}
\usepackage{listings}
\usepackage{amsfonts}
\usepackage{float}
\usepackage{tikzit}
\usetikzlibrary{decorations.pathreplacing}
\tikzstyle{new style 0}=[fill=white, draw=none, shape=rectangle]
\tikzstyle{new style 1}=[fill=white, draw=black, shape=circle]
\tikzstyle{new style 2}=[fill=white, draw=black, shape=rectangle]
\tikzstyle{new edge style 7}=[decorate, decoration={brace,amplitude=5pt, mirror}, xshift=-4pt, yshift=0pt]
\tikzstyle{new style 3}=[fill=white, draw=black, shape=circle, densely dotted]
\tikzstyle{new style 4}=[fill=white, draw=white, shape=circle]
\tikzstyle{new style 5}=[fill=white, draw=black, dotted, shape=circle]

\tikzstyle{new edge style 0}=[<-, draw=black, tikzit fill={rgb,255: red,25; green,255; blue,44}, tikzit draw=black]
\tikzstyle{new edge style 1}=[draw=black, {|->}, fill=white, tikzit draw=black]
\tikzstyle{new edge style 2}=[draw=none, fill=none, -, tikzit fill=none]
\tikzstyle{new edge style 3}=[draw={rgb,255: red,125; green,58; blue,71}, <-]
\tikzstyle{new edge style 4}=[<-, draw={rgb,255: red,35; green,66; blue,77}]
\tikzstyle{new edge style 5}=[dashed, <-]
\tikzstyle{new edge style 6}=[dashed, -, draw={rgb,255: red,193; green,193; blue,193}]

\usepackage{cite}
\usepackage{wrapfig}
\usepackage{sectsty}
\usepackage{titlesec}
\usepackage{comment}
\usepackage{ amssymb }
\usepackage{subcaption}
\usepackage[font=small]{caption}
\usepackage[colorlinks]{hyperref}
\hypersetup{
    colorlinks = true,
    linkcolor=black,
    citecolor=black,
    urlcolor=black
}
\definecolor{blue(ryb)}{rgb}{0.01, 0.28, 1.0}
\newcommand{\secref}[1]{\autoref{sec:#1}}
\newcommand{\figref}[1]{\autoref{fig:#1}}
\definecolor{bazaar}{rgb}{0.6, 0.47, 0.48}

\newcommand{\ascode}[1]{{\small\textsf{#1}}}

\newcommand{\getcurrentref}[1]{%
  \ifnum\value{#1}=0 ??\else\csname the#1\endcsname\fi
}
\newcommand{\textsfsmall}[1]{\scriptsize\textsf{#1}}
\titleformat{\subsubsection}[runin]
  {\normalfont\normalsize\bfseries}{\thesubsubsection}{0.5em}{}
\titleformat{\paragraph}[runin]
{\normalfont\normalsize\bfseries}{\theparagraph}{1em}{}

\makeatletter
\newcommand*\bigcdot{\mathpalette\bigcdot@{.5}}
\newcommand*\bigcdot@[2]{\mathbin{\vcenter{\hbox{\scalebox{#2}{$\m@th#1\bullet$}}}}}
\makeatother
\sectionfont{\large}
\subsectionfont{\normalsize}
\subsubsectionfont{\normalsize \normalfont \noindent \textit}
\paragraphfont{\normalfont \textit}

\begin{document}
\title{Folding over Neural Networks}
\author{
    Minh Nguyen
    \inst{1}
    \orcidID{0000-0003-3845-9928}
\and
    Nicolas Wu
    \inst{2}
    \orcidID{0000-0002-4161-985X}
}
    \institute{University of Bristol, Bristol, UK\\
    \email{min.nguyen@bristol.ac.uk}\\
\and
    Imperial College London, London, UK\\
    \email{n.wu@imperial.ac.uk}
}
\maketitle
\begin{abstract}
  % Functional programming has many useful mathematical abstractions to offer to the practice of deep learning; in particular
  Neural networks are typically represented as data structures that are traversed either through iteration or by manual chaining of method calls. However, a deeper analysis reveals that structured recursion can be used instead, so that traversal is directed by the structure of the network itself. This paper shows how such an approach can be realised in Haskell, by encoding neural networks as recursive data types, and then their training as recursion scheme patterns. In turn, we promote a coherent implementation of neural networks that delineates between their structure and semantics, allowing for compositionality in both how they are built and how they are trained.
\end{abstract}

%   In the past, neural networks have been typically represented using data structures that are traversed using iteration. Modern libraries take a more compositional approach that instead follows the structure of that data by using function calls specified by the user. However, a deeper analysis shows that this pattern can be made more explicit by allowing the data structure itself to direct the computation by using recursion schemes.

% In Keras, a Sequential Model has a layers field which is a list or stack of Layer objects. https://github.com/keras-team/keras/blob/master/keras/engine/sequential.py
% A Model class (which can be built using the functional api) does not have a layers field, but just an initial input and final output which are the results of layer composition. each layer is then connected to another, and calling one layer necessarily calls the previous layer. calling a model necessarily will invoke them all.
% https://github.com/keras-team/keras/blob/v2.8.0/keras/engine/training.py#L66-L3144
% https://github.com/keras-team/keras/blob/d8fcb9d4d4dad45080ecfdd575483653028f8eda/keras/engine/base_layer.py#L88

% recursion is driven by data itself rather than the explicit function calls instrumented by the programmer,
% we dont want to talk about instrumenting the recursion at all.


\section{Introduction}
\label{sec:intro}

Neural networks are graphs whose nodes and edges are organised into layers, generally forming a sequence of layers:
\begin{figure}[H]
  \vspace{-0.45cm}
  \centering
  % \vspace{1cm}
  \resizebox{0.65\textwidth}{!}{
    \centering\captionsetup[subfigure]{justification=centering}
    \begin{tikzpicture}
	\begin{pgfonlayer}{nodelayer}
		\node [style=new style 1] (0) at (-2.75, 0.75) {};
		\node [style=new style 1] (1) at (-1, 1.5) {};
		\node [style=new style 1] (2) at (-2.75, -0.75) {};
		\node [style=new style 1] (3) at (-1, 0) {};
		\node [style=new style 1] (4) at (-1, -1.5) {};
		\node [style=new style 1] (5) at (1, 1.5) {};
		\node [style=new style 1] (6) at (1, 0) {};
		\node [style=new style 1] (7) at (1, -1.5) {};
		\node [style=new style 1] (8) at (2.5, 0.75) {};
		\node [style=new style 1] (9) at (2.5, -0.75) {};
		\node [style=none] (10) at (-4.25, 0.75) {};
		\node [style=none] (11) at (-4.25, -0.75) {};
		\node [style=none] (12) at (3.75, 0.75) {};
		\node [style=none] (13) at (3.75, -0.75) {};
		\node [style=none] (14) at (-5.75, 0) {\scriptsize input};
		\node [style=none] (15) at (5.5, 0) {\scriptsize output};
		\node [style=none] (16) at (-4.5, -2.75) {};
		\node [style=none] (17) at (3.75, -2.75) {};
		\node [style=none] (18) at (4, 2.75) {};
		\node [style=none] (19) at (-4.5, 2.75) {};
		\node [style=none] (20) at (0, 3.5) {\scriptsize \textit{forward propagation}};
		\node [style=none] (21) at (0, -3.5) {\scriptsize \textit{back propagation}};
	\end{pgfonlayer}
	\begin{pgfonlayer}{edgelayer}
		\draw (5) to (1);
		\draw (6) to (3);
		\draw (7) to (4);
		\draw (1) to (0);
		\draw (3) to (0);
		\draw (3) to (2);
		\draw (4) to (2);
		\draw (1) to (2);
		\draw (4) to (0);
		\draw (5) to (3);
		\draw (6) to (4);
		\draw (6) to (1);
		\draw (7) to (3);
		\draw (5) to (4);
		\draw (7) to (1);
		\draw (8) to (5);
		\draw (9) to (6);
		\draw (9) to (7);
		\draw (8) to (6);
		\draw (9) to (5);
		\draw (8) to (7);
		\draw [style=new edge style 0] (0) to (10.center);
		\draw [style=new edge style 0] (2) to (11.center);
		\draw [style=new edge style 0] (12.center) to (8);
		\draw [style=new edge style 0] (13.center) to (9);
		\draw [style=new edge style 5] (16.center) to (17.center);
		\draw [style=new edge style 5] (18.center) to (19.center);
	\end{pgfonlayer}
\end{tikzpicture}
  }
  \vspace{-0.45cm}
\end{figure}
\noindent
Given input data (on the left), which is propagated through a series of transformations performed by each layer, the final output (on the right) assigns some particular meaning to the input. This process is called \emph{forward propagation}. To improve the accuracy of neural networks, their outputs are compared with expected values, and the discrepencies are sent back in the reverse direction through each layer to appropriately update their parameters. This is \emph{back propagation}.

How these notions are typically implemented is highly influenced by object-oriented and imperative programming design, where leading frameworks such as Keras \cite{keras} and PyTorch \cite{pytorch} can build on the extensive, existing support for machine learning in their host language. However, the design patterns of these paradigms tend to forgo certain appealing abstractions of neural networks; for example, one could perhaps view the diagram above as a composition of functions as layers, whose overall network structure is described by a higher-order function. Such concepts are more easily captured by functional languages, where networks can be represented as mathematical objects that are amenable to interpretation.

The relationship of neural networks with functional programming has been demonstrated several times, introducing support for compositional and type-safe implementation in various manners \cite{olah2015neural, elliott2018simple, grenade}. Using Haskell, this paper explores a categorical narrative that offers structure and compositionality in new ways: %we show in Haskell how neural networks can be abstracted into generic recursive data structures, and their operations of forward and back propagation into patterns of \emph{folds} and \emph{unfolds} over these structures.

% After giving the necessary background in \secref{background}, our contributions are as follows:
\begin{itemize}[leftmargin=2mm, label=\textbullet]
  \item We illustrate how fully connected networks can be expressed as fixed-points of recursive data structures, and their operations of forward and back propagation as patterns of \emph{folds} and \emph{unfolds} over these structures. Neural network training (forward then back propagation) is then realised as a composition of fold and unfold (\secref{fullyconnected}).
  \item We generalise our definition of neural networks into their types of layers by using coproducts, and provide an interface  for modularly constructing networks using free monads (\secref{freemonads}).
  \item We show how neural network training can be condensed into a single fold (\secref{singlefold}).
\end{itemize}

\noindent
We represent the ideas above with \textit{structured recursion schemes} \cite{hinze2013unifying}. By doing so, we create a separation of concern between what the layers of a neural network do from how they comprise the shape of the overall network; this then allows compositionality to be developed in each of these areas individually.

% Recursion schemes enjoy a number of universal properties \cite{meijer1991functional}, such as having unique solutions and being well-formed, making it possible to equationally reason about the construction and evaluation of data structures; the setting of structured recursion can also give rise to significant performance gains via fusion properties \cite{wu2015fusion}.

A vast number of neural network architectures are sequentially structured~\cite{bebis1994feed, wang2014generalized, hinton2009deep}. This paper hence uses fully connected networks \cite{svozil1997introduction} as a running example, being simple yet representative of this design. In turn, we believe the ideas presented are transferable to networks with more complex sequential structure, perhaps being sequential across multiple directions; we have tested this with convolutional networks \cite{yamashita2018convolutional} and recurrent networks \cite{medsker2001recurrent} in particular.

\section{Background}
\label{sec:background}
We begin by giving the necessary background to the recursion schemes used throughout this paper. First, consider the well-known recursive \lstinline{List} type:
\begin{lstlisting}
  data List a = Nil |$\vert$| Cons a (List a)
\end{lstlisting}
Folding and unfolding over a list are then defined as:

\noindent
\hspace{-0.35cm}
\begin{minipage}{.5\textwidth}
\begin{lstlisting}
foldr| :: |(a -> b -> b) -> b -> List a -> b
foldr f z Nil         = z
foldr f z (Cons x xs) = f x (foldr f z xs)
\end{lstlisting}
\end{minipage}% This must go next to `\end{minipage}`
\begin{minipage}{.7\textwidth}
  \begin{lstlisting}[]
    unfoldr| :: |(b -> Maybe (a, b)) -> b -> List a
    unfoldr f b = case f b of
      Just (a, b') -> Cons a (unfoldr f b')
      Nothing      -> Nil
    \end{lstlisting}
\end{minipage}
Here, \lstinline{foldr} recursively evaluates over a list of type \lstinline{[a]} to an output of type \lstinline{b}, by iteratively applying an accumulator function \lstinline{f} (with some base value \lstinline{z}). Conversely, \lstinline{unfoldr} recursively builds a list of type \lstinline{[a]} from a seed value \lstinline{b}, by iteratively applying some generator function \lstinline{g}.

%  by using \emph{recursion schemes} -- composable combinators that generalize over folds or unfolds \cite{hinze2013unifying}.

\subsubsection{Non-recursive functors and fix points}
The above useful patterns of recursion can be generalised to work over \textit{arbitrary} nested data types, by requiring a common structure to recurse over -- in particular, as a \emph{functor} \lstinline{f}:
\begin{lstlisting}
  class Functor f where
    fmap :: (k -> l) -> f k -> f l
\end{lstlisting}
Additionally, \ascode{f} is required to be non-recursive, and recursion is instead represented abstractly in its type parameter \lstinline{k} of \ascode{f k}. The function \lstinline{fmap} can then support generic mappings to the recursive structure captured by \lstinline{k}.

For example, the standard type \lstinline{List} can be converted into a non-recursive functor \ascode{\lstinline{ListF} a}:
\begin{lstlisting}
  data ListF a k = NilF |$\vert$| ConsF a k |\vspace{0.1cm}|
  instance Functor (ListF a) where
    fmap f NilF        = NilF
    fmap f (ConsF a k) = ConsF a (f k)
\end{lstlisting}
\noindent
which is functorial over the new type parameter \ascode{k}, implicitly representing the recursive occurrence of \lstinline{ListF}.

All explicit recursion is then instead relocated to the \emph{fixed-point} type \lstinline{Fix f}:
\begin{lstlisting}
  newtype Fix f = In (f (Fix f)) |\vspace{0.15cm}|
  out :: Functor f => Fix f -> f (Fix f)
  out (In f) = f
\end{lstlisting}
The constructor \ascode{In} wraps a recursive structure of type \ascode{f (\lstinline{Fix} f)} to yield the type \ascode{\lstinline{Fix} f}, and the function \ascode{out} deconstructs this to reattain the type \ascode{f (\lstinline{Fix} f)}.
Values of type \ascode{\lstinline{Fix} f} hence encode the generic recursive construction of a functor \lstinline{f}.

For example, the list \ascode{[1, 2]} in its \lstinline{Fix} form would be represented as:
\begin{lstlisting}
  |In (ConsF 1 (In (ConsF 2 (In NilF))))| :: Fix |(|ListF| |Int)
\end{lstlisting}
As \lstinline{NilF} contains no parameter \lstinline{k}, it encodes the base case (or least fixed-point) of the structure.

The ideas introduced so far provide us a setting for defining and using recursion schemes; this paper makes use of two in particular: catamorphisms (folds) and anamorphisms (unfolds).

\subsubsection{Catamorphisms }
A \emph{catamorphism}, given by the function \ascode{cata}, generalises over folds of lists to arbitrary algebraic data types \cite{malcolm1990data}.

\begin{lstlisting}
  cata :: Functor f => (f a -> a) -> Fix f -> a
  cata alg = alg . fmap (cata alg) . out
\end{lstlisting}
The argument \lstinline{alg} is called an \emph{f-algebra} or simply \emph{algebra}, being a function of type \lstinline{f a -> a}; this describes how a data structure of type \ascode{f a} is evaluated to an underlying value of type \ascode{a}. The type \ascode{a} is referred to as the algebra's \emph{carrier type}.

Informally, \ascode{cata} recursively evaluates a structure of type \lstinline{Fix f} down to an output of type \lstinline{a}, by  unwrapping the constructor of \lstinline{Fix} via \lstinline{out}, and then interpreting constructors of \lstinline{f} with \ascode{alg}.

% The function \ascode{cata alg} is total when \lstinline{alg} is total.

\subsubsection{Anamorphisms }
Conversely, an \emph{anamorphism}, given by \ascode{ana}, generalises over unfolds of lists to arbitrary algebraic data types \cite{meijer1991functional}.
\begin{lstlisting}
  ana :: Functor f => (b -> f b) -> b -> Fix f
  ana coalg = In . fmap (ana coalg) . coalg
\end{lstlisting}
The argument \ascode{coalg} is called an \emph{f-coalgebra} or simply \emph{coalgebra}, being a function of type \lstinline{b -> f b}; this describes how a structure of type \lstinline{f b} is constructed from an initial value of type \ascode{b}, where \ascode{b} is the carrier type of the coalgebra.

The function \lstinline{ana} recursively generates a structure of type \lstinline{Fix f} from a seed value of type \lstinline{b}, by using \ascode{coalg} to replace occurrences of \lstinline{b} with constructors of \lstinline{f}, and then wrapping the result with the \lstinline{Fix} constructor \lstinline{In}.

\noindent

\section{Fully Connected Networks}
\label{sec:fullyconnected}

We now consider how fully connected networks, one of the simplest types of neural networks, can be realised as an algebraic data type for structured recursion to operate over. These consist of a series of layers whose nodes are connected to all nodes in the previous and next layer:

\begin{figure}[H]
  \vspace{-0.3cm}
  \hspace{-1.5cm}
  \centering
  % \vspace{1cm}
  \resizebox{0.6\textwidth}{!}{
    \centering\captionsetup[subfigure]{justification=centering}
    \begin{tikzpicture}
	\begin{pgfonlayer}{nodelayer}
		\node [style=new style 1] (0) at (-3, 0.75) {};
		\node [style=new style 1] (1) at (-1, 1.5) {};
		\node [style=new style 1] (2) at (-3, -0.75) {};
		\node [style=new style 1] (3) at (-1, 0) {};
		\node [style=new style 1] (4) at (-1, -1.5) {};
		\node [style=new style 1] (5) at (1, 1.5) {};
		\node [style=new style 1] (6) at (1, 0) {};
		\node [style=new style 1] (7) at (1, -1.5) {};
		\node [style=new style 1] (8) at (2.75, 0.75) {};
		\node [style=new style 1] (9) at (2.75, -0.75) {};
		\node [style=none] (10) at (-4.5, 0.75) {};
		\node [style=none] (11) at (-4.5, -0.75) {};
		\node [style=none] (12) at (4, 0.75) {};
		\node [style=none] (13) at (4, -0.75) {};
		\node [style=none] (14) at (-3.75, -3.25) {\scriptsize input layer};
		\node [style=none] (16) at (0, -3.25) {\scriptsize dense layer};
		\node [style=none] (17) at (-1, -2.25) {};
		\node [style=none] (18) at (1, -2.25) {};
		\node [style=none] (22) at (-4.5, -2.25) {};
		\node [style=none] (23) at (-3, -2.25) {};
		\node [style=none] (24) at (-1.75, -2.25) {$\ldots$};
		\node [style=none] (25) at (1.75, -2.25) {$\ldots$};
		\node [style=none] (26) at (-2, 2.25) {\tiny $l = 1$};
		\node [style=none] (27) at (0, 2.25) {\tiny $l = 2$};
		\node [style=none] (28) at (2, 2.25) {\tiny $l = 3$};
		\node [style=none] (29) at (-4, 2.25) {\tiny $l = 0$};
		\node [style=none] (30) at (-6, 2.25) {\tiny layer $l$};
		\node [style=none] (31) at (-3, 2) {};
		\node [style=none] (32) at (-1, 2) {};
		\node [style=none] (33) at (1, 2) {};
		\node [style=none] (34) at (2.75, 2) {};
		\node [style=none] (35) at (2.75, -2.25) {};
	\end{pgfonlayer}
	\begin{pgfonlayer}{edgelayer}
		\draw (5) to (1);
		\draw (6) to (3);
		\draw (7) to (4);
		\draw (1) to (0);
		\draw (3) to (0);
		\draw (3) to (2);
		\draw (4) to (2);
		\draw (1) to (2);
		\draw (4) to (0);
		\draw (5) to (3);
		\draw (6) to (4);
		\draw (6) to (1);
		\draw (7) to (3);
		\draw (5) to (4);
		\draw (7) to (1);
		\draw (8) to (5);
		\draw (9) to (6);
		\draw (9) to (7);
		\draw (8) to (6);
		\draw (9) to (5);
		\draw (8) to (7);
		\draw [style=new edge style 0] (0) to (10.center);
		\draw [style=new edge style 0] (2) to (11.center);
		\draw [style=new edge style 0] (12.center) to (8);
		\draw [style=new edge style 0] (13.center) to (9);
		\draw [style=new edge style 7] (17.center) to (18.center);
		\draw [style=new edge style 7] (22.center) to (23.center);
		\draw [style=new edge style 6] (31.center) to (23.center);
		\draw [style=new edge style 6] (32.center) to (17.center);
		\draw [style=new edge style 6] (33.center) to (18.center);
		\draw [style=new edge style 6] (34.center) to (35.center);
	\end{pgfonlayer}
\end{tikzpicture}
  }
  \vspace{-0.5cm}
\end{figure}

\noindent
The functor \ascode{f} chosen to represent this structure is the layer type, \lstinline{Layer}:

\begin{lstlisting}
  data Layer k  = |InputLayer| Values |$\vert$| |DenseLayer| Weights Biases k deriving Functor |\vspace{0.15cm}|
  type Values   = [Double]
  type Biases   = [Double]
  type Weights  = [[Double]]
\end{lstlisting}

The case \ascode{InputLayer} is the first layer of the network and contains only the network's initial input \lstinline{Values}. This is the base case of the functor.

The case \ascode{DenseLayer} is \emph{any subsequent layer} (including the output layer), and contains as parameters a matrix \lstinline{Weights} and vector \lstinline{Biases} which are later used to transform a given input. Its argument \lstinline{k} then represents its previous connected layer as a recursive parameter.

Notice that the dimensions of \lstinline{Weights} and \lstinline{Biases} in fact sufficiently describe a layer's internal structure:

\begin{figure}[H]
  \vspace{-0.5cm}
  \centering
  % \vspace{1cm}
  \resizebox{0.6\textwidth}{!}{
    \centering\captionsetup[subfigure]{justification=centering}
    \begin{tikzpicture}
	\begin{pgfonlayer}{nodelayer}
		\node [style=new style 5] (1) at (-3.75, 1) {};
		\node [style=new style 5] (3) at (-3.75, -1.25) {};
		\node [style=new style 1] (5) at (0.5, 2) {};
		\node [style=new style 1] (6) at (0.5, 0) {};
		\node [style=new style 1] (7) at (0.5, -2) {};
		\node [style=none] (16) at (-1.75, -2.75) {\scriptsize $l^{th}$ dense layer };
		\node [style=none] (26) at (0.5, 2.75) {\scriptsize $b_l^0$};
		\node [style=none] (27) at (0.5, 0.75) {\scriptsize $b_l^1$};
		\node [style=none] (29) at (-2, 2) {\scriptsize $w_l^{(0,0)}$};
		\node [style=none] (30) at (-2, 0.75) {\scriptsize $w_l^{(1,0)}$};
		\node [style=none] (32) at (5.75, -0.25) {\scriptsize $[[w_l^{(0, 0)}, w_l^{(1, 0)}],$};
		\node [style=none] (33) at (5.75, -1.25) {\scriptsize $\;[w_l^{(0, 1)}, w_l^{(1, 1)}],$};
		\node [style=none] (35) at (2.75, -0.25) {\scriptsize $w_l =$  };
		\node [style=none] (36) at (2.75, 0.75) {\scriptsize $b_l =$  };
		\node [style=none] (37) at (5, 0.75) {\scriptsize $[\,b_l^0, b_l^1, b_l^2]$};
		\node [style=none] (38) at (0.5, -1.25) {\scriptsize $b_l^2$};
		\node [style=none] (39) at (5.75, -2.25) {\scriptsize $\;[w_l^{(0, 2)}, w_l^{(1, 2)}]]$};
	\end{pgfonlayer}
	\begin{pgfonlayer}{edgelayer}
		\draw (5) to (1);
		\draw [style=new edge style 6] (6) to (3);
		\draw [style=new edge style 6] (7) to (3);
		\draw (3) to (5);
		\draw [style=new edge style 6] (1) to (6);
		\draw [style=new edge style 6] (1) to (7);
	\end{pgfonlayer}
\end{tikzpicture}
  }
  \vspace{-0.5cm}
\end{figure}
\noindent
In the example layer $l$ above, each $j^{th}$ node has a bias value $b_l^j$, and the edge to the $j^{th}$ node from the previous layer's $i^{th}$ node has a weight value $w_l^{(i,j)}$. Hence, a weight matrix with dimensions $n \times m$ specifies $n$ nodes in the current layer, with each node having $m$ in-degrees. One could make this structure explicit by choosing a more precise type such as vectors \cite{grenade}, but we avoid this for simplicity.

By then incorporating \lstinline{Fix}, an instance of a network is represented as a recursive nesting of layers of type \lstinline{Fix Layer}. For example, below corresponds to the fully connected network shown at the beginning of \secref{fullyconnected}:

\begin{lstlisting}
  fixNetwork :: Fix Layer
  fixNetwork = In (|DenseLayer| |$w_3$| |$b_3$|           -- |$l=3, \; \textsf{dims}(w_3) = 2\times3$|
                (In (|DenseLayer| |$w_2$| |$b_2$|         -- |$l=2, \; \textsf{dims}(w_2) = 3\times3$|
                  (In (|DenseLayer| |$w_1$| |$b_1$|       -- |$l=1, \; \textsf{dims}(w_1) = 3\times2$|
                    (In (|InputLayer| |$a_0$|)))))) |\,|  -- |$l=0, \; \textsf{dims}(a_0) \,= 2$|
\end{lstlisting}
% \begin{lstlisting}
%   fixNetwork :: Fix Layer
%   fixNetwork = In (|DenseLayer| |$w^3$| |$b^3$|                 -- fourth layer
%                   (In (|DenseLayer| |$w^2$| |$b^2$|             -- third layer
%                       (In (|DenseLayer| |$w^1$| |$b^1$|         -- second layer
%                           (In (|InputLayer| |$a^0$|))))))    |\,|-- input layer
% \end{lstlisting}
A comparison can be drawn between the above construction and that of lists, where \ascode{DenseLayer} and \ascode{InputLayer} are analogous to \lstinline{Cons} and \lstinline{Nil}. Of course, there are less arduous ways of constructing such values, and a monadic interface for doing this is later detailed in \secref{freemonads}.

The type \lstinline{Fix Layer} then provides a base for encoding the operations of neural networks -- forward and back propagation -- as recursion schemes.

\subsection{Forward propagation as a catamorphism}
\label{sec:forward-prop}

The numerous end-user applications that neural networks are well-known for, such as facial recognition \cite{khashman2009application} and semantic parsing \cite{yih2014semantic}, are all done via \emph{forward propagation}: given unstructured input data, this is propagated through a sequence of transformations performed by each network layer; the final output can then be interpreted meaningfully by humans, for example, as a particular decision or some classification of the input.

One may observe that forward propagation resembles that of a fold: given an input, the layers of a neural network are recursively collapsed and evaluated to an output value (analogous to folding a list whose elements are layers). We can implement this notion using a generalised fold -- a \emph{catamorphism} -- over the type \lstinline{Fix Layer}; to do so simply requires a suitable algebra to fold with.

\vspace{-0.1cm}
\paragraph{An algebra for forward propagation} An algebra, \lstinline{f a -> a}, specialised to our context will represent forward propagation over a single layer. The functor \lstinline{f} is hence \lstinline{Layer}. The choice of carrier type \lstinline{a} is \lstinline{[Values]}, that is, the accumulation of outputs of all previous layers. This gives rise to the following type signature:
\begin{lstlisting}
  alg|$_{\textsfsmall{fwd}}$| :: Layer [Values] -> [Values]
\end{lstlisting}

\noindent Defining the case for \ascode{InputLayer} is trivial: a singleton list containing only the initial input $a_0$ is passed forward to the next layer.

\begin{lstlisting}
  alg|$_{\textsfsmall{fwd}}$| (|InputLayer| |$a_0$|)  = |$[a_0]$|
\end{lstlisting}

\noindent Defining the case for \ascode{DenseLayer} is where any numerical computation is involved:

\begin{lstlisting}
  alg|$_{\textsfsmall{fwd}}$| (|DenseLayer| |$w_{l}$ $b_{\,l}$| (|$a_{l-1}\,$:$\,as$|)) = (|$a_{l}$\,|:|\,$a_{l-1}\,$:$\,as$|)
    where |$a_{l}$| = |$\sigma(w_{l} *  a_{l-1} + b_{l})$| |\vspace{0.15cm}|
\end{lstlisting}

\noindent
Above, the output $a_{l-1}$ of the previous layer is used as input for the current $l^{th}$ layer, letting the next output $a_l$ be computed; this is given by multiplying $a_{l-1}$ with weights $w_{l}$, adding on biases $b_{l}$, and applying some normalization function $\sigma$ (we assume the correct operators $*$ and $+$ for matrices or vectors, given fully in \secref{apdx:forward-prop}). The output $a_{l}$ is then prepended to the list of previous outputs.

Having defined forward propagation over a single layer, recursively performing this over an entire neural network is done by applying \ascode{cata alg$_{\textsfsmall{fwd}}$} to a value of type \lstinline{Fix Layer}, yielding each of its layers' outputs: \vspace{0.05cm}

\begin{lstlisting}
  cata alg|$_{\textsfsmall{fwd}}$| :: Fix Layer -> [Values]
\end{lstlisting}
This decoupling of non-recursive logic (\ascode{alg$_{\textsfsmall{fwd}}$}) from recursive logic (\lstinline{cata}) provides a concise description of how a layer transforms its input to its output, without concerning the rest of the network structure. %The separation between the syntax and semantics of networks also makes it straightforward to implement other behaviours for forward propagation as algebras; for instance, one could imagine also tracing execution data for debugging, or  augmenting the algorithm with new computational steps.

\subsubsection{A better algebra for forward propagation} The initial input to a neural network is currently stored as an argument of the \ascode{InputLayer} constructor:
\begin{lstlisting}
  -- currently
  data Layer k = |InputLayer| Values |$\vert$| |DenseLayer| Weights Biases k
\end{lstlisting}
However, this design is somewhat simplistic. Rather, a neural network should be able to exist independently of its input value like so:

% Therefore, let us first change the data type for |Layer| by removing the input parameter
% from the constructor |InputLayer|.

\begin{lstlisting}
  data Layer k = |InputLayer| |$\vert$| |DenseLayer| Weights Biases k
\end{lstlisting}
and have its input be provided externally instead. To implement this, we will have \ascode{alg$_{\textsfsmall{fwd}}$} evaluate a layer to a \emph{continuation} of type \ascode{\lstinline{Values} \lstinline{->}\,[\lstinline{Values}]} that awaits an input before performing forward propagation:
\begin{lstlisting}
  alg|$_{\textsfsmall{fwd}}$| :: Layer (Values -> [Values]) -> (Values -> [Values])
\end{lstlisting}

In the case of \ascode{InputLayer}, this returns a function that wraps some provided initial input into a singleton list:
\begin{lstlisting}
  alg|$_{\textsfsmall{fwd}}$| |InputLayer| = |$\lambda a_0$| -> [|$a_0$]|
\end{lstlisting}

For \ascode{DenseLayer}, its argument ``\ascode{forwardPass}'' is the continuation built from forward propagating over the previous layers:
\begin{lstlisting}
  alg|$_{\textsfsmall{fwd}}$| (|DenseLayer| |$w_{l}$ $b_{\,l}$| forwardPass)
    = (|$\lambda$|(|$a_{l-1}\,$:$\,as$|) -> let |$a_{l}$| = |$\sigma(w_{l} * a_{l-1} + b_{l})$| in (|$a_{l}$|:|$a_{l-1}\,$:$\,as$|)) . forwardPass
\end{lstlisting}
This is composed with a new function that takes the previous outputs \ascode{($a_{l-1}\,$:$\,as$)} and prepends the current layer's output $a_l$, as defined before.

Folding over a neural network with the above algebra will then return the composition of each layer's forward propagation function:

\begin{lstlisting}
  cata alg|$_{\textsfsmall{fwd}}$| :: Fix Layer -> (Values -> [Values])
\end{lstlisting}
Given an initial input, the type \ascode{\lstinline{Values} \lstinline{->}\,[\lstinline{Values}]} returns a list of all the layers' resulting outputs.

% \vspace{-0.1cm}
\subsection{Back propagation as an anamorphism}
\label{sec:back-prop-ana}
Using a neural network to extract meaning from input data, via forward propagation above, is only useful if the network produces accurate outputs in the first place; this is determined by the ``correctness'' of its  \lstinline{Weights} and \lstinline{Biases} parameters. Learning these parameters is called \emph{back propagation}: given the \emph{actual} output of forward propagation, it proceeds in the reverse direction of the network by updating each layer's parameters  with respect to a \emph{desired} output.

One could hence view back propagation as resembling an unfold, which recursively constructs an updated neural network from right to left. Dually to \secref{forward-prop}, we can encode this as a generalised unfold -- an \emph{anamorphism} -- over the type \lstinline{Fix Layer}; to do so simply requires an appropriate coalgebra to unfold with.

\paragraph{A coalgebra for back propagation} A coalgebra, \lstinline{b -> f b}, will represent back propagation over a single layer. As before, \lstinline{f} is \lstinline{Layer}. The choice of carrier type \lstinline{b} is slightly involved, as it should denote the information to be passed backwards through each layer, letting their weights and biases be correctly updated.

In particular, to update the $l^{th}$ layer requires knowledge of:

\hspace{-1cm}
\begin{minipage}{.59\textwidth}
\begin{enumerate}[leftmargin=*, label=(\roman*)]
  \setlength\itemsep{0.7em}
  % \item We need the desired final network output with respect to the initial input.
  \item Its original input $a_{l-1}$ and output $a_l$. \label{item:input-stack}
  \item The next layer's weights $w_{l+1}$ and \emph{delta value} $\delta_{l+1}$, the latter being the output error of that layer.  If there is no next layer, the \emph{desired output} of the entire network is needed instead.
\end{enumerate}

\end{minipage}% This must go next to `\end{minipage}`
\begin{minipage}{.5\textwidth}
\begin{figure}[H]
  \vspace{-0.5cm}
  \centering
  % \vspace{1cm}
  \resizebox{1\textwidth}{!}{
    \centering\captionsetup[subfigure]{justification=centering}
    \begin{tikzpicture}
	\begin{pgfonlayer}{nodelayer}
		\node [style=new style 5] (1) at (-1, 1.5) {};
		\node [style=new style 5] (3) at (-1, 0) {};
		\node [style=new style 5] (4) at (-1, -1.5) {};
		\node [style=new style 1] (5) at (1, 1.5) {};
		\node [style=new style 1] (6) at (1, 0) {};
		\node [style=new style 1] (7) at (1, -1.5) {};
		\node [style=none] (17) at (0.25, -3.25) {};
		\node [style=none] (18) at (2, -3.25) {};
		\node [style=none] (27) at (-1.25, 2.5) {{\tiny \textsf{input}} \scriptsize $a_{l-1}$};
		\node [style=none] (28) at (1.25, 2.5) {{\tiny \textsf{output}} \scriptsize $a_l$};
		\node [style=none] (29) at (2.5, -2.75) {{\tiny \textsf{error}} \scriptsize $\delta_{l+1}$};
		\node [style=none] (30) at (2.25, -2.25) {{\tiny  \textsf{weights}} \scriptsize $w_{l+1}$};
		\node [style=none] (31) at (-1, 2) {};
		\node [style=none] (32) at (0, 2) {};
		\node [style=none] (33) at (1, 2) {};
		\node [style=none] (34) at (2, 2) {};
		\node [style=new style 5] (35) at (3, 1.5) {};
		\node [style=new style 5] (36) at (3, 0) {};
		\node [style=new style 5] (37) at (3, -1.5) {};
		\node [style=none] (39) at (5, -2.25) {\tiny  \textsf{desired} };
		\node [style=none] (40) at (5, -2.75) {\tiny  \textsf{output}};
		\node [style=none] (42) at (4, -2.5) {\tiny  \emph{or} };
	\end{pgfonlayer}
	\begin{pgfonlayer}{edgelayer}
		\draw (5) to (1);
		\draw (6) to (3);
		\draw (7) to (4);
		\draw (5) to (3);
		\draw (6) to (4);
		\draw (6) to (1);
		\draw (7) to (3);
		\draw (5) to (4);
		\draw (7) to (1);
		\draw [style=new edge style 0] (17.center) to (18.center);
		\draw [style=new edge style 0] (32.center) to (31.center);
		\draw [style=new edge style 0] (34.center) to (33.center);
		\draw [style=new edge style 6] (5) to (36);
		\draw [style=new edge style 6] (5) to (35);
		\draw [style=new edge style 6] (37) to (5);
		\draw [style=new edge style 6] (6) to (36);
		\draw [style=new edge style 6] (7) to (37);
		\draw [style=new edge style 6] (7) to (36);
		\draw [style=new edge style 6] (6) to (37);
		\draw [style=new edge style 6] (6) to (35);
	\end{pgfonlayer}
\end{tikzpicture}
  }
\end{figure}
\end{minipage}

\vspace{0.3cm}
\noindent
This is all captured by the type \lstinline{BackProp} below, where $as$ is the list of all layers' inputs and outputs produced from forward propagation:
\begin{lstlisting}
type Deltas    = [Double]
data BackProp = |BackProp| { |$as$|            |\,\,::| [Values]
                           , |$w_{l+1}$|         |\,::| Weights
                           , |$\delta_{l+1}$|          |::| Deltas
                           , desiredOutput |::| Values }
\end{lstlisting}
\noindent
The choice of carrier type \lstinline{b}, in coalgebra \lstinline{b -> Layer b}, is then \lstinline{(Fix Layer, BackProp)}:
\begin{lstlisting}
  coalg|$_{\textsfsmall{bwd}}$| :: (Fix Layer, BackProp) |$\rightarrow$| Layer (Fix Layer, BackProp)
\end{lstlisting}
As well as containing the information that is passed back through each layer, it also contains the \emph{original network} of type \lstinline{Fix Layer}. Defining \ascode{coalg$_{\textsfsmall{bwd}}$} thus consists of pattern matching against values of \lstinline{Fix Layer}, determining the network structure to be generated.

When matching against \ascode{In InputLayer}, there are no parameters to update and so \ascode{InputLayer} is trivially returned:
\begin{lstlisting}
  coalg|$_{\textsfsmall{bwd}}$| (In (|InputLayer|), |$\underline{\;\;}$|) = |InputLayer|
\end{lstlisting}

When matching against \ascode{In DenseLayer}, the layer's output error and updated parameters must be computed:
\begin{lstlisting}
  coalg|$_{\textsfsmall{bwd}}$| (In (|DenseLayer| |$w_l$| |$b_l$| prevLayer), backProp|$_{l+1}$|)
    = let (|$\delta_{l}$|, |$w^{new}_{l}$|, |$b^{new}_{l}$|) = backward |$w_l$| |$b_l$| backProp|$_{l+1}$|
          backProp|$_{l}$| = |BackProp| { |$w_{l+1}$| = |$w_l$|,
                                  |$\delta_{l+1}$| |\,|= |$\delta_{l}$|,
                                  |$as$|   = tail (|$as$| backProp|$_{l+1}$|) }
      in |DenseLayer| |$w^{new}_{l}$| |$b^{new}_{l}$| (prevLayer, backProp|$_{l}$|) |\vspace{0.25cm}|
  backward :: Weights -> Biases -> BackProp -> (Deltas, Weights, Biases)
\end{lstlisting}
Above assumes the auxiliary function \lstinline{backward} (given fully in \secref{apdx:back-prop}): this takes as arguments the old parameters $w_l$ and $b_l$, and the back propagation values \ascode{backProp$_{l+1}$} computed by the \emph{next} layer; it then returns an output error $\delta_{l}$ and updated weights $w^{new}_{l}$ and biases $b^{new}_{l}$.

Next, \ascode{backProp$_{l}$} constructs the data to be passed to the \emph{previous} layer: this stores the old weights, the newly computed delta, and the tail of the outputs $as$; the last point ensures the head of $as$ is always the original output of the layer being updated. Finally, a new \ascode{DenseLayer} is returned with updated parameters.

Having implemented back propagation over a single layer, recursively updating an entire network, \lstinline{Fix Layer}, is then done by calling \ascode{ana coalg$_{\textsfsmall{bwd}}$} (provided some initial value of type \lstinline{BackProp}):
\begin{lstlisting}
  ana coalg|$_{\textsfsmall{bwd}}$| :: (Fix Layer, BackProp) -> Fix Layer
\end{lstlisting}
This can be incorporated alongside forward propagation to define the more complete procedure of neural network \emph{training}, as shown next.

\subsection{Training neural networks with metamorphisms}

Transforming an input through a neural network to an output (forward propagation), and then optimising the network parameters according to a desired output (back propagation), is known as \emph{training}; the iteration of this process prepares a network to be reliably used for real-world applications.

Training can hence be viewed as the composition of forward and back propagation, which in our setting, is a catamorphism (fold) followed by an anamorphism (unfold), also known as a \textit{metamorphism} \cite{gibbons2004streaming}. We encode this below as \ascode{train}: given an initial input $a_0$, a corresponding desired output, and a neural network \ascode{nn}, this performs a single network update:
\begin{lstlisting}
  train :: (Values, Values) -> Fix Layer -> Fix Layer
  train (|$a_0$|, desiredOutput) nn = (ana |coalg$_{\textsfsmall{bwd}}$| . h . cata |alg$_{\textsfsmall{fwd}}$|) nn
      where h :: (Values -> [Values]) -> (Fix Layer, BackProp)
           |\,| h forwardPass = let |$as$| = forwardPass |$a_0$|
                           |\,\,| in (nn, |BackProp| |$as$| [] [] desiredOutput)
\end{lstlisting}

First, forward propagation is performed by applying \ascode{cata alg$_{\textsfsmall{fwd}}$} to \ascode{nn}, producing a function of type \ascode{\lstinline{Values}}\,$\rightarrow$ \ascode{[\lstinline{Values}]}.

The intermediary function \ascode{h} then maps the \emph{output} of forward propagation, \ascode{forwardPass :: \lstinline{Values}}\,$\rightarrow$ \ascode{[\lstinline{Values}]}, to the \emph{input} of back propagation, which has type \lstinline{(Fix Layer, BackProp)}. Here, \ascode{forwardPass} is applied to the initial input \ascode{$a_0$} to yield the outputs $as$ of all layers, which are returned alongside the desired output and original network \ascode{nn}.

Lastly, back propagation is performed by \ascode{ana coalg$_{\textsfsmall{bwd}}$}. From the seed value of type \lstinline{(Fix Layer, BackProp)}, it generates a network with new weights and biases.

Updating a neural network over many inputs and desired outputs is then simple, and can be defined by folding with \ascode{train}; this is shown in \secref{training-example}.

\section{Neural Networks \`a la Carte}
\label{sec:freemonads}

Below shows how one would use the previous implementation to represent a fully connected neural network, consisting of an input layer and two dense layers.
\begin{lstlisting}
  fixNetwork :: Fix Layer
  fixNetwork = In (|DenseLayer| |$w_2$| |$b_2$| (In (|DenseLayer| |$w_1$| |$b_1$| (In |InputLayer|))))
\end{lstlisting}
\noindent
Such values of type \lstinline{Fix Layer} can be rather cumbersome to write, and require all of their layers to be declared non-independently at the same time. A further orthogonal issue is that each kind of layer, \ascode{DenseLayer} and \ascode{InputLayer}, exclusively belongs to the type \lstinline{Layer} of fully connected networks; in reality, the same kinds of layers are commonly reused in many different network designs, but the current embedding does not support modularity in this way.

We resolve these matters by taking influence from the \emph{data types \`a la carte} approach \cite{swierstra2008data}, and incorporate free monads and coproducts in our recursion schemes. The result allows neural networks to be defined modularly as coproducts of their types of layers, and then constructed using monadic \ascode{do}-notation:

\begin{lstlisting}
  freeNetwork :: Free (InputLayer :+: DenseLayer) a
  freeNetwork = do
    denselayer |$w_2$| |$b_2$|
    denselayer |$w_1$| |$b_1$|
    inputlayer
\end{lstlisting}
This then enables sections of networks to be independently defined in terms of the layers they make use of, and then later connected:

\hspace{-0.3cm}
\begin{minipage}{.3\textwidth}
\begin{lstlisting}[]
freeNetwork = do
  network|$_2$|
  network|$_1$|
\end{lstlisting}
\end{minipage}% This must go next to `\end{minipage}`
\begin{minipage}{.7\textwidth}
\begin{lstlisting}[]
network|$_2$| :: DenseLayer :<: |\,|f => Free f ()
network|$_2$| = do denselayer |$w_2$| |$b_2$|
\end{lstlisting}
\begin{lstlisting}[]
network|$_1$| :: (|$\!$|InputLayer :<: |\,|f, DenseLayer :<:|\,|f) => Free f a
network|$_1$| = do denselayer |$w_1$| |$b_1$|
              inputlayer
\end{lstlisting}
\end{minipage}

\vspace{-0.3cm}
\subsection{Free monads and coproducts}
\label{ssec:freemonads}

\subsubsection{Free monads} The type \lstinline{Fix f} currently forces neural networks to be declared in one go. Free monads, of the type \lstinline{Free f a}, instead provide a monadic interface for writing elegant, composable constructions of functors \lstinline{f}:

\begin{lstlisting}
--     Fix  f   = In (f (Fix f))
  data Free f a = Op (f (Free f a)) |$\vert$| Pure a |\vspace{0.15cm}|
  instance Functor f => Monad (Free f) where
    return a     |\,|= Pure a
    Pure a >>= |\,| k = k a
    Op f   >>= |\,| k = Op (fmap (>>=) k)
\end{lstlisting}
Identical to \lstinline{In} of \lstinline{Fix f}, the constructor \lstinline{Op} of \lstinline{Free f a} also encodes the generic recursion of \lstinline{f}; the constructor \lstinline{Pure} then represents a return value of type \lstinline{a}. The key property of interest, is that monadically binding with \texttt{(>>=)} in the free monad corresponds to extending its recursive structure at the most nested level.

% this will allow segments of networks to be independently define and connected.

Using this, a fully connected network would have type \lstinline{Free Layer a}, and an example of one input layer and two dense layers can be constructed like so:
\begin{lstlisting}
  freeNetwork|$'$| :: Free Layer a
  freeNetwork|$'$| = do
   Op (|DenseLayer| |$w_2$| |$b_2$| (Pure ()))
   Op (|DenseLayer| |$w_1$| |$b_1$| (Pure ()))
   Op |InputLayer|
\end{lstlisting}

\vspace{-0.3cm}
\subsubsection{Coproducts} To then promote modularity in the kinds of layers, the constructors of \lstinline{Layer} are redefined with their own types:
\begin{lstlisting}
  data DenseLayer a = |DenseLayer| Weights Biases a deriving Functor
  data InputLayer a = |InputLayer| deriving Functor
\end{lstlisting}
A type that contains these two layers can be described using the \emph{coproduct} type \lstinline{f :+: g}, being the coproduct of two functors \lstinline{f} and \lstinline{g}:
\begin{lstlisting}
  data (f :+: g) a = L (f a) |$\vert$| R (g a) deriving Functor
\end{lstlisting}
For example, a fully connected network would correspond to the free monad whose functor is the coproduct \lstinline{InputLayer :+: DenseLayer}:

\begin{lstlisting}
  type FullyConnectedNetwork a = Free (InputLayer :+: DenseLayer) a
\end{lstlisting}
This supports reusability of layers when defining different variations of networks. As a simple example, one could represent a convolutional network by extending \lstinline{FullyConnectedNetwork} with a convolutional layer (\secref{convolutional}):
\begin{lstlisting}
  type ConvolutionalNetwork a = Free (InputLayer :+: DenseLayer :+: ConvLayer) a
\end{lstlisting}

To then instantiate and pattern match against coproduct values, the type class \lstinline{sub :<:  sup} is used, stating that \lstinline{sup} is a type signature that contains \lstinline{sub}.
\begin{lstlisting}
  class (Functor sub, Functor sup) => sub :<: sup where
    inj :: sub a -> sup a
    prj :: sup a -> Maybe (sub a)
\end{lstlisting}
When this constraint holds, there must be a way of \emph{injecting} a value of type \ascode{sub a} into \lstinline{sup a}, and a way of \emph{projecting} from \lstinline{sup a} back into a value of type \lstinline{Maybe (sub a)}. These can be used to define ``smart constructors'' for each layer, as seen before in \ascode{freeNetwork}:

\begin{lstlisting}
  denselayer :: (DenseLayer :<:  f) => Weights -> Biases -> Free f ()
  denselayer |$w$| |$b$| = Op (inj (|DenseLayer| |$w$| |$b$| (Pure ())))

  inputlayer :: (InputLayer :<:  f) =>  Free f a
  inputlayer = Op (inj |InputLayer|)
\end{lstlisting}
The above provides an abstraction for injecting layers into the type \lstinline{Free f} where \lstinline{f} is a coproduct of those layers.

\subsection{Training neural networks with free monads}

We next turn to reimplementing the recursion schemes \lstinline{cata} and \lstinline{ana} to support free monads and coproducts -- this then involves minor revisions to how the algebra and coalgebra for forward and back propagation are represented.

\subsubsection{Forward propagation with free monads}
\label{ssec:freecata}
The free monadic version of \lstinline{cata} is known as \lstinline{eval}, and evaluates a structure of \lstinline{Free f a} to a value of type \lstinline{b}:
\begin{lstlisting}
  eval :: Functor f => (f b -> b) -> (a -> b) -> Free f a -> b
  eval alg gen (Pure x) = gen x
  eval alg gen (Op f)   = (alg . fmap (eval alg gen)) f
\end{lstlisting}
Above carries out the same procedure as \lstinline{cata}, but also makes use of a \emph{generator} \lstinline{gen :: a -> b}; the generator's purpose is to interpret values of type \lstinline{a} in \lstinline{Free f a} into a desired carrier type \lstinline{b} for \ascode{alg :: f b $\rightarrow$ b}. We remark that \lstinline{eval} can in fact be defined in terms of \lstinline{cata}, and is thus also a catamorphism. %(We note that as \lstinline{InputLayer} has no arguments, our concrete representation of networks never contain the \ascode{Pure} constructor, and so \lstinline{gen})

Some minor changes are then made to the forward propagation algebra: as each layer now has its own type, defining \ascode{alg$_{\textsfsmall{fwd}}$} for each of these requires ad-hoc polymorphism. The type class \lstinline{AlgFwd f} is hence introduced which layers \ascode{f} can derive from to support forward propagation:

\begin{lstlisting}
  class Functor f => AlgFwd f where
    |alg$_{\textsfsmall{fwd}}$| :: f (Values -> [Values]) -> (Values -> [Values])
\end{lstlisting}

\noindent
The algebra \ascode{alg$_{\textsfsmall{fwd}}$} for constructors \ascode{InputLayer} and \ascode{DenseLayer} are unchanged:

\begin{lstlisting}
  instance AlgFwd InputLayer where
    alg|$_{\textsfsmall{fwd}}$| |InputLayer|  = |$\lambda a_0$| -> [|$a_0$|]
  instance AlgFwd DenseLayer where
    alg|$_{\textsfsmall{fwd}}$| (|DenseLayer| |$w_{l}$ $b_{\,l}$| forwardPass)
        = (|$\lambda$|(|$a_{l-1}\,$:$\,as$|) -> let |$a_{l}$| = |$\sigma(w_{l} * a_{l-1} + b_{l})$| in (|$a_{l}$|:|$a_{l-1}\,$:$\,as$|)) . forwardPass
\end{lstlisting}

\noindent
Using  \ascode{alg$_{\textsfsmall{fwd}}$} with \lstinline{eval} is then similar to as with \lstinline{cata}:
\begin{lstlisting}
  eval alg|$_{\textsfsmall{fwd}}$| gen|$_{\textsfsmall{fwd}}$| :: AlgFwd f => Free f a -> (Values -> [Values])
    where gen|$_{\textsfsmall{fwd}}$| :: a -> (Values -> [Values])
          |\,|gen|$_{\textsfsmall{fwd}}$| = const (|$\lambda$|x -> [x])
\end{lstlisting}
This recursively evaluates a neural network of type \lstinline{Free f a} to yield its forward propagation function. Here, the generator \ascode{gen$_{\textsfsmall{fwd}}$} is chosen as \ascode{const ($\lambda$x $\rightarrow$ [x])}, mapping the type \lstinline{a} in \lstinline{Free f a} to the desired carrier \lstinline{Values -> [Values]}.

\subsubsection{Back propagation with free monads} The free monadic version of \lstinline{ana} is known as \lstinline{build} (which can also be defined in terms of \ascode{ana}), and generates a structure of \lstinline{Free f a} from a seed value of type \lstinline{b}:
\begin{lstlisting}
  build :: Functor f => (b -> f b) -> b -> Free f a
  build coalg = Op .  fmap (ana f) . coalg
\end{lstlisting}

Similar to \lstinline{AlgFwd f}, the type class \lstinline{CoalgBwd f} is defined from which derived instances \lstinline{f} implement back propagation via its \ascode{coalg$_{\textsfsmall{bwd}}$} method:
\begin{lstlisting}
  class Functor f => CoalgBwd f where
    coalg|$_{\textsfsmall{bwd}}$| :: (Free f a, BackProp) -> f (Free f a, BackProp)
\end{lstlisting}
Here, the coalgebra's carrier now has the neural network to be updated as the type \lstinline{Free f} rather than \lstinline{Fix f}. Also note that unlike \lstinline{AlgFwd f} whose instances correspond to \emph{individual layers}, the instances of \lstinline{f} derived for \lstinline{CoalgBwd f} correspond to the \emph{entire network} to be constructed.

For fully connected networks,  \lstinline{f} would be \lstinline{(InputLayer :+: DenseLayer)}:

\begin{lstlisting}
  instance CoalgBwd (InputLayer :+: DenseLayer) where
\end{lstlisting}

Defining \ascode{coalg$_{\textsfsmall{bwd}}$} is then the familiar process of pattern matching on the structure of the original network (\secref{back-prop-ana}), determining the updated network to be generated. Rather than directly matching on values of \lstinline{(InputLayer :+: DenseLayer) a}, we make use of \emph{pattern synonyms} \cite{wu2014effect} for a less arduous experience:
\begin{lstlisting}
  pattern |InputLayer'| <- (prj -> Just |InputLayer|)
  pattern |DenseLayer'| |$w_l$| |$b_l$| prevLayer <- (prj -> Just (|DenseLayer| |$w_l$| |$b_l$| prevLayer))
\end{lstlisting}
Here, matching a layer against the pattern synonym \ascode{InputLayer$'$} would be equivalent to it passing it to \lstinline{prj} below:
\begin{lstlisting}
  prj :: InputLayer :<:  f => f a -> Maybe (InputLayer a)
\end{lstlisting}
and then successfully pattern matching against \ascode{Just InputLayer}. A similar case holds for \ascode{DenseLayer'}.

The coalgebra \ascode{coalg$_{\textsfsmall{bwd}}$} for \ascode{InputLayer'} and \ascode{DenseLayer'} are then the same as with \lstinline{Fix}, but now the updated layers are injected into a coproduct:
\begin{lstlisting}
  coalg|$_{\textsfsmall{bwd}}$| (Op |InputLayer'|, |$\underline{\;\;}$|) = inj |InputLayer|
  coalg|$_{\textsfsmall{bwd}}$| (Op (|DenseLayer'| |$w_l$| |$b_l$| prevLayer), backProp|$_{l+1}$|) =
    let (|$w^{new}_{l}$|, |$b^{new}_{l}$|, |$\delta_{l}$|) = |$\ldots$|
        backProp|$_{l}$|         |\,|= |$\ldots$|
    in  inj (|DenseLayer| |$w^{new}_{l}$| |$b^{new}_{l}$| (prevLayer, backProp|$_l$|))
\end{lstlisting}

\noindent
Using \ascode{coalg$_{\textsfsmall{bwd}}$} with \ascode{build} instead of \ascode{ana} is also familiar, but now generates neural networks of the type \lstinline{Free f}:

\begin{lstlisting}
  build coalg|$_{\textsfsmall{bwd}}$| :: AlgBwd f =>  (Free f a, BackProp) -> Free f b
\end{lstlisting}

\subsubsection{Training with free monads} Finally, \lstinline{eval} and \lstinline{build} can be combined to redefine the function \ascode{train}, now incorporating free monads and coproducts:
\begin{lstlisting}
train :: (AlgFwd f, CoalgBwd f) => (Values, Values) -> Free f a -> Free f a
train (|$a_0$|, desiredOutput) nn = (build coalg|$_{\textsfsmall{bwd}}$| . h . eval alg|$_{\textsfsmall{fwd}}$| gen|$_{\textsfsmall{fwd}}$|) nn
\end{lstlisting}
This maps a network of type \ascode{\lstinline{Free} f a}, where \ascode{f} implements forward and back propagation, to an updated network; the intermediary function \ascode{h} used is unchanged.

\section{Training with just Folds}
\label{sec:singlefold}

Currently, we encode forward then back propagation as a fold followed by an unfold. Rather than needlessly traversing the structure of a neural network {twice}, it is desirable to combine this work under a \emph{single traversal}. This can be made possible by matching the recursion patterns used for each propagation. We show this by encoding back propagation as a \emph{fold} instead (\secref{backcata}), which can then be performed alongside forward propagation under a single fold (\secref{singlefold-ssec}).

\subsection{Back propagation as a fold}
\label{sec:backcata}

Back propagation begins from the end of a neural network and progresses towards the start, the reason being that each layer is updated with respect to the next layer's computations. An anamorphism may seem like a natural choice of recursive pattern for this: it  generates the outermost constructor (final layer) first, and then recurses by generating nested constructors (thus progressing towards the input layer). This is in contrast to \ascode{cata} which begins evaluating from the most nested constructor and then progresses outwards.

However, just as \ascode{unfold} can be defined in terms of \ascode{fold}, \ascode{coalg$_{\textsfsmall{bwd}}$} can be redefined as a back propagation \emph{algebra} \ascode{alg$_{\textsfsmall{bwd}}$} for folding with:

\begin{lstlisting}
  class AlgBwd g f where
    alg|$_{\textsfsmall{bwd}}$| :: g (BackProp -> Free f a) -> (BackProp -> Free f a)
\end{lstlisting}
\noindent
In \lstinline{AlgBwd g f}, the parameter \ascode{g} is the specific layer that is back propagated over, and \ascode{f} is the entire network to be constructed. The algebra \ascode{alg$_{\textsfsmall{bwd}}$} then has as its carrier type the continuation \lstinline{BackProp -> Free f a}, which given back propagation values from the next layer, will update the current and  all previous layers.

% We also note that as \ascode{alg$_{\textsfsmall{bwd}}$} is applied directly to a layer \ascode{g}, it needs not carry the ``original'' neural network in its carrier type, in contrast to \ascode{coalg$_{\textsfsmall{bwd}}$}:
% \begin{lstlisting}
%   -- here, Free f a represented the original neural network to be updated
%   |\emph{coalg$_{\textsfsmall{bwd}}$ :: (Free f a, BackProp) $\rightarrow$ f (Free f a, BackProp)}|
% \end{lstlisting}

Defining \ascode{alg$_{\textsfsmall{bwd}}$} for \ascode{InputLayer} returns a function that always
produces an \ascode{InputLayer} (injected into the free monad).

\begin{lstlisting}
  instance InputLayer :<:  f => AlgBwd InputLayer f where
    alg|$_{\textsfsmall{bwd}}$| |InputLayer| = |const (inject InputLayer)| |\vspace{0.2cm}|
  inject| = |Op . inj
\end{lstlisting}

Defining \ascode{alg$_{\textsfsmall{bwd}}$} for \ascode{DenseLayer} returns a function that  uses the \ascode{backProp$_{l+1}$} value from the next layer to first update the current layer.

\begin{lstlisting}
  instance DenseLayer :<:  f => AlgBwd DenseLayer f where
    alg|$_{\textsfsmall{bwd}}$| (|DenseLayer| |$w_l$| |$b_{l}$| backwardPass) = |$\lambda$| backProp|$_{l+1}$| ->
      let (|$w^{new}_{l}$|, |$b^{new}_{l}$|, |$\delta_{l}$|) = |$\ldots$|
          backProp|$_{l}$|         |\,|= |$\ldots$|
      in  |inject (DenseLayer $w^{new}_{l}$ $b^{new}_{l}$ (backwardPass backProp$_{l}$))|
\end{lstlisting}
The computed \ascode{backProp$_{l}$} value is then passed to the continuation ``\ascode{backwardPass}'' of type \lstinline{BackProp -> Free f a}, invoking back propagation on all the \emph{previous} layers.

Folding back propagation over an entire network, via \ascode{eval alg$_{\textsfsmall{bwd}}$}, will then compose each layer's \lstinline{BackProp -> Free f a} function, resulting in a chain of promises between layers to update their previous layer:

\begin{lstlisting}
  eval alg|$_{\textsfsmall{bwd}}$| gen|$_{\textsfsmall{bwd}}$|| :: |(AlgBwd f f, InputLayer :<:  f) => Free f a -> (BackProp -> Free f a)
    where gen|$_{\textsfsmall{bwd}}$| :: a -> (BackProp -> Free f a)
          |\,|gen|$_{\textsfsmall{bwd}}$| |$\underline{\;\;}$| = |const (inject InputLayer)|
\end{lstlisting}
Here, the constraint \lstinline{AlgBwd f f} says we can evaluate a network \lstinline{f} to return a function that updates that same network. The generator \ascode{gen$_{\textsfsmall{bwd}}$} simply interprets the type \lstinline{a} of \lstinline{Free f a} to the desired carrier.

% We could now define the training of a neural network, \ascode{train}, as the composition of two catamorphisms -- one for \ascode{alg$_{\textsfsmall{fwd}}$}, and one for \ascode{alg$_{\textsfsmall{bwd}}$}. However, we need not traverse over the same network structure twice in $O(2n)$ time: instead, we can encode training as a \textit{single} catamorphism in $O(n)$.

\subsection{Training as a single fold}
\label{sec:singlefold-ssec}
Having implemented forward and back propagation as algebras, it becomes possible to perform both in the same fold; this circumstance is given rise to by the `banana split' property of folds \cite{meijer1991functional}, stating that any pair of folds over the same structure can always be combined into a single fold that generates a pair.

% To demonstrate, we will define an algebra that simultaneously computes the forwards and backwards pass functions over a network.

To do this, the function \ascode{pairGen} is defined to take two generators of types \ascode{b} and \ascode{c}, and return a generator for \ascode{(b, c)}. Similarly, the function \ascode{pairAlg} takes two algebras with carrier types \lstinline{b} and \lstinline{c}, and returns an algebra of carrier type \lstinline{(b, c)}.

\begin{lstlisting}
  pairGen :: (a -> b) -> (a -> c) -> a -> (b, c)
  pairGen gen|$_b$| gen|$_c$| a = (gen|$_b$| a, gen|$_c$| a)

  pairAlg :: Functor f => (f b -> b) -> (f c -> c) -> f (b, c) -> (b, c)
  pairAlg alg|$_b$| alg|$_c$| f|$_{bc}$| = (alg|$_b$| (fmap fst f|$_{bc}$|), alg|$_c$| (fmap snd f|$_{bc}$|))
\end{lstlisting}

These are incorporated below to redefine the function \ascode{train}, by first pairing the algebras and generators for forward and back propagation:

\begin{lstlisting}
  |\,train :: |(InputLayer :<:|\,| f, AlgFwd f, AlgBwd f f) => (Values, Values) -> Free f a -> Free f a
  train (|a$_0$|, desiredOutput) nn =
    let alg|$_{\textsfsmall{train}}$| = pairAlg alg|$_{\textsfsmall{fwd}}$| alg|$_{\textsfsmall{bwd}}$|
        gen|$_{\textsfsmall{train}}$| = pairGen gen|$_{\textsfsmall{fwd}}$| gen|$_{\textsfsmall{bwd}}$|
\end{lstlisting}
Folding over a neural network with these produces a function \lstinline{forwardPass} of type \lstinline{Values -> [Values]} and a function \lstinline{backwardPass} of type \lstinline{BackProp -> Free f a}:
\begin{lstlisting}
        (forwardPass, backwardPass) = eval alg|$_{\textsfsmall{train}}$| gen|$_{\textsfsmall{train}}$| nn
\end{lstlisting}
The intermediary function \ascode{h} is then defined to take the future output of \lstinline{forwardPass} and initialise a \lstinline{BackProp} value as input for \ascode{backwardPass}:
\begin{lstlisting}
        h :: [Values] -> BackProp
        h |$as$| = |BackProp| |$as$| [] [] desiredOutput
\end{lstlisting}
Finally, passing an initial input $a_0$ to the composition of \ascode{forwardPass}, \ascode{h}, and \ascode{backwardPass}, returns an updated network:
\begin{lstlisting}
    in  (backwardPass .|\,| h . forwardPass) |$a_0$|
\end{lstlisting}

\subsection{Example: training a fully connected network}
\label{sec:training-example}

We now show that our construction of fully connected networks is capable of learning. As an example, we will aim to model the sine function, such that the desired output of the network is the sine of a provided input value.

The network we construct is seen below on the right, where a single input value is propagated through three intermediate layers, and then the output layer produces a single value. Its implementation is given by \ascode{fcNetwork}, where \ascode{randMat2D m n} represents a randomly valued matrix of dimension \ascode{m$\times$n}.

\hspace{-1cm}
\begin{minipage}{.65\textwidth}
\begin{lstlisting}
  type FullyConnectedNetwork = (InputLayer :+: DenseLayer)

  fcNetwork :: Free FullyConnectedNetwork a
  fcNetwork = do
      denselayer (randMat2D 1 3) (randVec 1)  -- |$l_4$|
      denselayer (randMat2D 3 3) (randVec 3)  -- |$l_3$|
      denselayer (randMat2D 3 3) (randVec 3)  -- |$l_2$|
      denselayer (randMat2D 3 1) (randVec 3)  -- |$l_1$|
      inputlayer                             |\,\,\,|-- |$l_0$|
\end{lstlisting}
\end{minipage}% This must go next to `\end{minipage}`
\hspace{0.0cm}
\begin{minipage}{.5\textwidth}
\begin{figure}[H]
  % \vspace{-0.1cm}
  \centering
  % \vspace{1cm}
  \resizebox{0.9\textwidth}{!}{
    \centering\captionsetup[subfigure]{justification=centering}
    \begin{tikzpicture}
	\begin{pgfonlayer}{nodelayer}
		\node [style=new style 1] (0) at (-2.75, 0) {};
		\node [style=new style 1] (1) at (-1.5, 1.5) {};
		\node [style=new style 1] (3) at (-1.5, 0) {};
		\node [style=new style 1] (4) at (-1.5, -1.5) {};
		\node [style=new style 1] (5) at (0, 1.5) {};
		\node [style=new style 1] (6) at (0, 0) {};
		\node [style=new style 1] (7) at (0, -1.5) {};
		\node [style=none] (10) at (-4.25, 0) {};
		\node [style=new style 1] (22) at (1.5, 1.5) {};
		\node [style=new style 1] (23) at (1.5, 0) {};
		\node [style=new style 1] (24) at (1.5, -1.5) {};
		\node [style=new style 1] (25) at (2.75, 0) {};
		\node [style=none] (26) at (4, 0) {};
		\node [style=none] (30) at (-3.5, -2.5) {\scriptsize $l_0$};
		\node [style=none] (31) at (-2.25, -2.5) {\scriptsize $l_1$};
		\node [style=none] (32) at (-0.75, -2.5) {\scriptsize $l_2$};
		\node [style=none] (33) at (0.75, -2.5) {\scriptsize $l_3$};
		\node [style=none] (34) at (2, -2.5) {\scriptsize $l_4$};
		\node [style=none] (35) at (-2.75, 2) {};
		\node [style=none] (36) at (-2.75, -2.75) {};
		\node [style=none] (37) at (-1.5, 2) {};
		\node [style=none] (38) at (0, 2) {};
		\node [style=none] (39) at (1.5, 2) {};
		\node [style=none] (40) at (2.75, 2) {};
		\node [style=none] (41) at (-1.5, -2.75) {};
		\node [style=none] (42) at (0, -2.75) {};
		\node [style=none] (43) at (1.5, -2.75) {};
		\node [style=none] (44) at (2.75, -2.75) {};
	\end{pgfonlayer}
	\begin{pgfonlayer}{edgelayer}
		\draw (5) to (1);
		\draw (6) to (3);
		\draw (7) to (4);
		\draw (1) to (0);
		\draw (3) to (0);
		\draw (4) to (0);
		\draw (5) to (3);
		\draw (6) to (4);
		\draw (6) to (1);
		\draw (7) to (3);
		\draw (5) to (4);
		\draw (7) to (1);
		\draw [style=new edge style 0] (0) to (10.center);
		\draw (5) to (22);
		\draw (5) to (23);
		\draw (6) to (22);
		\draw (22) to (7);
		\draw (24) to (6);
		\draw (7) to (23);
		\draw (7) to (24);
		\draw (6) to (23);
		\draw (22) to (25);
		\draw (23) to (25);
		\draw (24) to (25);
		\draw [style=new edge style 0] (26.center) to (25);
		\draw [style=new edge style 6] (35.center) to (36.center);
		\draw [style=new edge style 6] (37.center) to (41.center);
		\draw [style=new edge style 6] (42.center) to (38.center);
		\draw [style=new edge style 6] (39.center) to (43.center);
		\draw [style=new edge style 6] (40.center) to (44.center);
	\end{pgfonlayer}
\end{tikzpicture}
  }
  \vspace{-0cm}
\end{figure}
\end{minipage}
\vspace{0.1cm}

\begin{figure}[H]
  % \centering
  % \vspace{-0.1cm}
  % \hspace*{-0.4cm}
  \begin{subfigure}{0.49\textwidth}
    \centering
    \includegraphics[width=1\columnwidth]{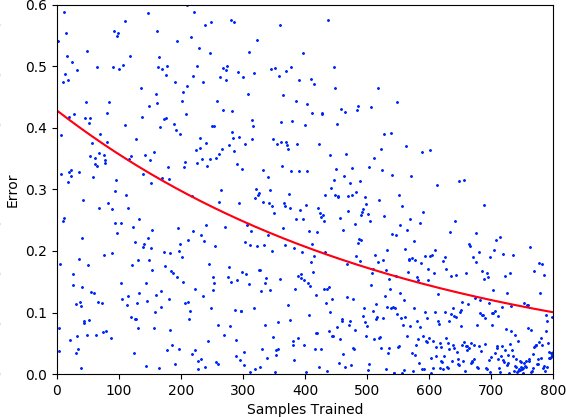}
    \vspace{-0.3cm}
  \end{subfigure}
  % \hspace*{0.1cm}
  \begin{subfigure}{0.49\textwidth}
    \centering
    \includegraphics[width=1\columnwidth]{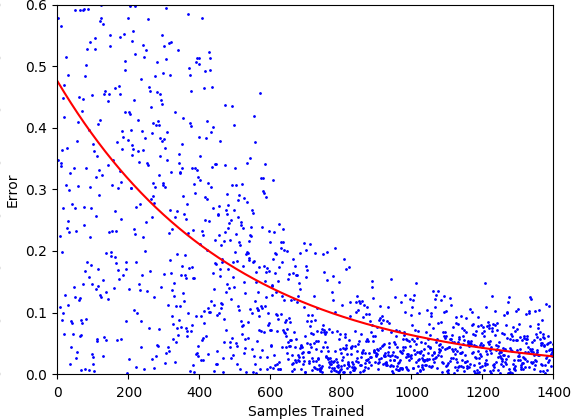}
    \vspace{-0.3cm}
    % \label{fig:fc_1400}
  \end{subfigure}
  \vspace{-0.1cm}
  \begin{subfigure}{0.49\textwidth}
    \centering
    \caption{Samples: 800, correlation coefficient: -0.54}
    \label{fig:fc_800}
  \end{subfigure}
  \begin{subfigure}{0.49\textwidth}
    \centering
    \caption{Samples: 1400, correlation coefficient: -0.65}
    \label{fig:fc_1400}
  \end{subfigure}
  \vspace{-0.2cm}
\caption{Training a fully connnected neural network}
\label{fig:fc_results}
\vspace{-0.3cm}
\end{figure}

To perform consecutive updates to the network over many inputs and desired outputs, one can define this as a straightforward list fold:
\begin{lstlisting}
  trainMany :: (InputLayer :<: |\,|f, AlgFwd f, AlgBwd f f)
            => [(Values, Values|$\!$|)] ->  Free f a ->  Free f a
  trainMany dataset nn = foldr train nn dataset
\end{lstlisting}

\noindent
An example program using this to train the network is given below:

\begin{lstlisting}
main n_samples = do
  g <- getStdGen
  let| initialInputs|    |\,|= take n_samples (randoms g)
      desiredOutputs = map sine| |initialInputs
      nn             = |trainMany (zip initialInputs desiredOutputs) fcNetwork|
\end{lstlisting}

\noindent
\figref{fc_results} shows the change in the output error of the neural network as more samples are trained with, comparing total sample sizes of 800 and 1400.  The error produced when classifying samples can be seen to gradually converge, where the negative curvature denotes the network's ability to progressively produce more accurate values. The higher magnitude in correlation coefficient on the right indicates a stronger overall rate of learning for the larger sample size.

\noindent

\vspace{-0.1cm}
\section{Summary}

This paper discussed a new narrative that offers structure and generality in describing neural networks, showing how they can be represented as recursive data structures, and their operations of forward and back propagation as recursion schemes. We demonstrated modularity in three particular ways: a delineation between the structure and semantics of a neural network, and compositionality in both neural network training and the types of layers in a network. Although only fully connected networks have been considered, we believe the ideas shown are transferrable to network types with more complex, sequential structure across multiple directions, and we have explored this externally with convolutional (\secref{convolutional}) and recurrent (\secref{lstm}) networks in particular.

There are a number of interesting directions that are beyond the scope of this paper. One of these is performance, in particular, establishing to what extent we trade off computational efficiency in order to achieve generality in our approach. It is hoped that some of this can be offset by the performance gains of ``fusion properties'' \cite{wu2015fusion} which the setting of structured recursion may give rise to; further insight would be needed as to the necessary circumstance for our implementation to exploit this.
A second direction is to explore the many useful universal properties that recursion schemes enjoy, such as having unique solutions and being well-formed \cite{meijer1991functional}; these may offer a means for equationally reasoning about the construction and evaluation of neural networks.

\subsection{Related work}

% \textit{Neural networks, types, and functional programming}
It has been long established that it is typically difficult to state and prove laws for arbitrary, explicitly recursive functions, and that structured recursion \cite{malcolm1990data, uustalu1999primitive, vene1998functional} can instead provide a setting where properties of termination and equational reasoning can be readily applied to programs \cite{meijer1991functional}. Neural networks are not typically implemented using recursion, and are instead widely represented as directed graphs of computation calls \cite{guresen2011definition}; the story between neural networks and recursion schemes hence has limited existing work. A particularly relevant discussion is presented by Olah \cite{olah2015neural} who corresponds the representation of neural networks to type theory in functional programming. This makes a number of useful intuitions, such as networks as chains of composed functions, and tree nets as catamorphisms and anamorphisms.

% Deep learning in general has taken steps towards more functional approaches, such as in Tensorflow \cite{tensorflow} and PyTorch \cite{pytorch}, where layers are objects which can be ``composed'' by nesting method calls to other layers.
Neural networks have seen previous exploration in Haskell. Campbell \cite{grenade} uses dependent types to implement recurrent neural networks, enabling type-safe composition of network layers of different shapes; here, networks are represented as heterogeneous lists of layers, and forward and back propagation over layer types as standard type class methods.

There is also work in the functional programming community on structured approaches towards graph types. Erwig \cite{erwig1997functional} takes a compositional view of graph, and defines graph algorithms in terms of folds that can facilitate program transformations and optimizations via fusion properties.  Oliveira and Cook \cite{oliveira2012functional} use parametric higher-order abstract syntax (PHOAS), and develop a language for representing structured graphs generically using fix points; operations on these graphs are then implemented as generalized folds.

Lastly, a categorical approach to deep learning has also been explored by Elliot \cite{elliott2018simple}, in particular towards automatic differentiation which is central in computing gradients during back propagation. They realise an intersection between category theory and computing derivatives of functions, and present a generalisation of automatic differentiation by replacing derivative values with an arbitrary cartesian category.

\titleformat{\section}{\large\bfseries}{\appendixname~\thesection .}{0.5em}{}

\bibliographystyle{splncs04}
\bibliography{bibliography}
\newpage
\appendix
\section{Elaborated code}
\vspace{-0.1cm}
\subsection{Forward propagation}
\label{sec:apdx:forward-prop}
Below gives the full implementation of \ascode{alg$_{\textsfsmall{fwd}}$} in \secref{forward-prop} for forward propagation.

\hspace{-0.4cm}
\begin{minipage}{0.6\textwidth}
\begin{lstlisting}
type Values = [Double]|\vspace{0.15cm}|
alg|$_{\textsfsmall{fwd}}$| :: Layer [Values] -> [Values]
alg|$_{\textsfsmall{fwd}}$| (|DenseLayer| |$w_{l}$ $b_{\,l}$| (|$a_{l-1}\,$:$\,as$|))
  = (|$a_{l}$\,|:|\,$a_{l-1}\,$:$\,as$|)
  where |$a_{l}$| = |$\sigma(w_{l} *  a_{l-1} + b_{l})$| |\vspace{0.15cm}|
-- vector addition
|$\oplus_{\tiny\textsf{v}}$| :: [Double] -> [Double] -> [Double]
xs |$\oplus_{\tiny\textsf{v}}$| ys  = zipWith (|+|) xs ys |\vspace{0.1cm}|
-- matrix-vector multiplication
|$\otimes_{\tiny\textsf{mv}}$| :: [[Double]] -> [Double] -> [Double]
xss |$\otimes_{\tiny\textsf{mv}}$| ys = map (sum . |\,| zipWith (*) ys) yss |\vspace{0.1cm}|
-- sigmoid function
|$\sigma$| :: [Double] -> [Double]
|$\sigma$| = map (|$\lambda x$| -> |$1 / (1 + e^{-x})$|)
\end{lstlisting}
\end{minipage}
\begin{minipage}{0.5\textwidth}
\begin{figure}[H]
  \vspace{-0.8cm}
  \centering
  % \vspace{1cm}
  \resizebox{1\textwidth}{!}{
    \centering\captionsetup[subfigure]{justification=centering}
    \begin{tikzpicture}
	\begin{pgfonlayer}{nodelayer}
		\node [style=new style 1] (1) at (-4.75, 3.5) {\scriptsize \color{white}{$a^0_{l\,}\;$}};
		\node [style=new style 4] (5) at (3, 1.25) {\scriptsize \color{white}{$a_l^0$}};
		\node [style=none] (26) at (3, 3.75) {\normalsize $+\; b_{\,l}^{\,0}$};
		\node [style=none] (41) at (3, 1.25) {\normalsize $\sum$};
		\node [style=new style 1] (42) at (-4.75, -1.25) {\color{white}{\scriptsize $a^1_{l\,}\;$}};
		\node [style=new style 4] (48) at (-1.25, 2.5) {\quad\quad\quad};
		\node [style=none] (49) at (-1, 2.5) {\normalsize $* \; w_l^{(0,0)}$};
		\node [style=new style 4] (51) at (-1.25, 0) {\quad\quad\quad};
		\node [style=none] (52) at (-1, 0) {\normalsize $* \; w_l^{(1,0)}$};
		\node [style=none] (53) at (0.25, 2) {};
		\node [style=none] (54) at (0, 0.25) {};
		\node [style=none] (56) at (-4.75, -3) {Input};
		\node [style=none] (58) at (8, -3.25) {Output};
		\node [style=none] (59) at (3, 3.25) {};
		\node [style=none] (60) at (-4.75, -1.25) {\normalsize $a^1_{l-1}$};
		\node [style=none] (61) at (-4.75, 3.5) {\normalsize $a^0_{l-1}$};
		\node [style=new style 1] (62) at (8, 1.25) {\normalsize{$a_l^0$}};
		\node [style=new style 4] (64) at (5.5, 1.25) {\normalsize $\sigma$};
	\end{pgfonlayer}
	\begin{pgfonlayer}{edgelayer}
		\draw [style=new edge style 0] (48) to (1);
		\draw [style=new edge style 0] (51) to (42);
		\draw [style=new edge style 0] (5) to (53.center);
		\draw [style=new edge style 0] (5) to (54.center);
		\draw [style=new edge style 0] (5) to (59.center);
		\draw [style=new edge style 0] (64) to (5);
		\draw [style=new edge style 0] (62) to (64);
	\end{pgfonlayer}
\end{tikzpicture}
  }
  \vspace{-0.3cm}
\end{figure}
\end{minipage}

\subsection{Back propagation}
\label{sec:apdx:back-prop}
\noindent
Below gives the full implementation of \ascode{backward}, used during back propagation.
\begin{lstlisting}
 backward :: Weights -> Biases -> BackProp -> (Deltas, Weights, Biases)
 backward |$w_l$| |$b_l$| (BackProp |$(a_l : a_{l-1} : as) \; w_{l+1} \; \delta_{l+1}$| desiredOutput) =
    let |$\delta_{l}$|    = case |$\delta_{l+1}$| of
                  [|\,|] -> (|$a_l \, \ominus_{\tiny\textsf{v}} $| desiredOutput)      |$\,\,\otimes_{\tiny\textsf{v}} \; \sigma'(a_{l-1})\;$| |\,|-- compute |$\delta_{l}$| for final layer
                  |$\underline{\;\;}$|  -> ((transpose |$w_{l + 1}) \; \otimes_{\tiny\textsf{mv}} \; \delta_{l+1}) \, \otimes_{\tiny\textsf{v}} \, \, \sigma'(a_{l-1})$|  -- compute |$\delta_{l}$| for any other layer
        |$w_l^{new}$| = |$w_l \ominus_{\tiny\textsf{m}} (a_{l-1} \circledcirc \delta_l )$|
        |$b_l^{new}$\,\,| = |$b_l \ominus_{\tiny\textsf{v}} \delta_l $|
    in (|$\delta_{l}$|, |$w^{new}_{l}$|, |$b^{new}_{l}$|)|\vspace{0.15cm}|
  -- vector multiplication
 |$\otimes_{\tiny\textsf{v}}$| :: [Double] -> [Double] -> [Double]
 xs |$\otimes_{\tiny\textsf{v}}$| ys  = zipWith (*) xs ys |\vspace{0.1cm}|
 -- vector subtraction
 |$\ominus_{\tiny\textsf{v}}$| :: [Double] -> [Double] -> [Double]
 xs |$\ominus_{\tiny\textsf{v}}$| ys  = zipWith (-) xs ys |\vspace{0.1cm}|
 -- matrix subtraction
 |$\ominus_{\tiny\textsf{m}}$| :: [[Double]] -> [[Double]] -> [[Double]]
 xss |$\ominus_{\tiny\textsf{m}}$| yss  = zipWith (|$\ominus_{\tiny\textsf{v}}$|) xs ys |\vspace{0.1cm}|
 -- outer product
 |$\circledcirc$| :: [Double] -> [Double] -> [[Double]]
 xs |$\circledcirc$| ys = map (|$\lambda$|x -> map (x *) ys) xs |\vspace{0.1cm}|
 -- inverse then differential sigmoid function
 |$\sigma'$| :: [Double] -> [Double]
 |$\sigma'$| = map (|$\lambda$|x -> let y = log(x/(1 - x)) in y * (1 - y))
\end{lstlisting}

\section{Training convolutional and recurrent networks}
\subsection{Training a convolutional neural network}
\label{sec:convolutional}

Convolutional neural networks are used primarily to classify images. To demonstrate the ability of a convolutional network implementation to learn, we choose to classify matrix values as corresponding to either an image displaying the symbol `\textsf{X}' or an image displaying the symbol `\textsf{O}'.

\begin{figure}[H]
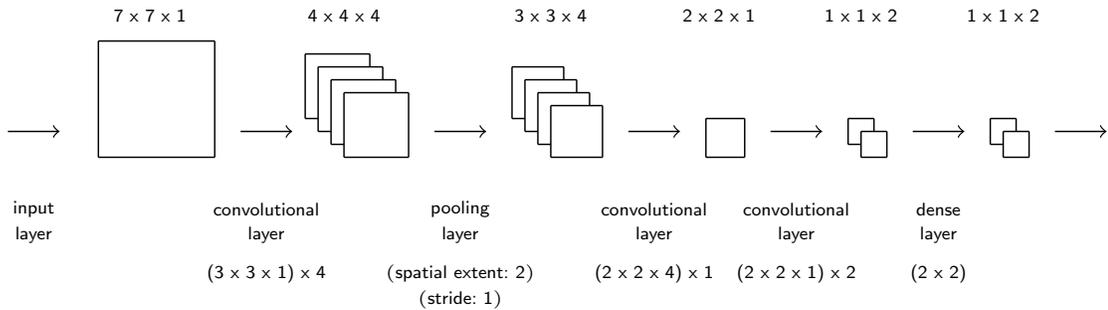

  \caption{Convolutional network}
  \label{fig:conv_network_appendix}
  \vspace{-1cm}
\end{figure}

A diagram of the neural network used can be seen in \figref{conv_network_appendix} where dimensions are given in the form {(width} $\times$ {height} $\times$ {depth)} $\times$ {number of filters}. An input image of dimensions ($7 \times 7 \times 1$) is provided to the network, resulting in an output vector of length two
where each value corresponds to the probability of the image being classified as either an `\textsf{X}' or an `\textsf{O}' symbol.

\hspace{-1.8cm}
\begin{minipage}{\textwidth}
\begin{figure}[H]
  \vspace{-0.1cm}
  \centering
  % \vspace{1cm}
  \resizebox{1.25\textwidth}{!}{
    \centering\captionsetup[subfigure]{justification=centering}
    \begin{tikzpicture}
	\begin{pgfonlayer}{nodelayer}
		\node [style=none] (0) at (-7.5, 3.25) {};
		\node [style=none] (1) at (-5.25, 3.25) {};
		\node [style=none] (2) at (-7.5, 1) {};
		\node [style=none] (3) at (-5.25, 1) {};
		\node [style=none] (4) at (-2.75, 2.25) {};
		\node [style=none] (5) at (-1.5, 2.25) {};
		\node [style=none] (6) at (-2.75, 1) {};
		\node [style=none] (7) at (-1.5, 1) {};
		\node [style=none] (8) at (-1.75, 2.5) {};
		\node [style=none] (9) at (-3, 2.5) {};
		\node [style=none] (10) at (-3, 1.25) {};
		\node [style=none] (11) at (-2.75, 1.25) {};
		\node [style=none] (12) at (-1.75, 2.25) {};
		\node [style=none] (13) at (-3.25, 2.75) {};
		\node [style=none] (14) at (-3.25, 1.5) {};
		\node [style=none] (15) at (-2, 2.75) {};
		\node [style=none] (16) at (-2, 2.5) {};
		\node [style=none] (17) at (-3, 1.5) {};
		\node [style=none] (19) at (-3.5, 3) {};
		\node [style=none] (20) at (-3.5, 1.75) {};
		\node [style=none] (21) at (-3.25, 1.75) {};
		\node [style=none] (22) at (-2.25, 3) {};
		\node [style=none] (23) at (-2.25, 2.75) {};
		\node [style=none] (24) at (-2.75, 3.75) {\tiny \textsf{4 x 4 x 4}};
		\node [style=none] (25) at (1.25, 2) {};
		\node [style=none] (26) at (2.25, 2) {};
		\node [style=none] (27) at (1.25, 1) {};
		\node [style=none] (28) at (2.25, 1) {};
		\node [style=none] (29) at (2, 2.25) {};
		\node [style=none] (30) at (1, 2.25) {};
		\node [style=none] (31) at (1, 1.25) {};
		\node [style=none] (32) at (1.25, 1.25) {};
		\node [style=none] (33) at (2, 2) {};
		\node [style=none] (34) at (0.75, 2.5) {};
		\node [style=none] (35) at (0.75, 1.5) {};
		\node [style=none] (36) at (1.75, 2.5) {};
		\node [style=none] (37) at (1.75, 2.25) {};
		\node [style=none] (38) at (1, 1.5) {};
		\node [style=none] (39) at (0.5, 2.75) {};
		\node [style=none] (40) at (0.5, 1.75) {};
		\node [style=none] (41) at (0.75, 1.75) {};
		\node [style=none] (42) at (1.5, 2.75) {};
		\node [style=none] (43) at (1.5, 2.5) {};
		\node [style=none] (44) at (7.25, 1.5) {};
		\node [style=none] (45) at (7.75, 1.5) {};
		\node [style=none] (46) at (7.25, 1) {};
		\node [style=none] (47) at (7.75, 1) {};
		\node [style=none] (49) at (4.25, 1.75) {};
		\node [style=none] (50) at (5, 1.75) {};
		\node [style=none] (51) at (4.25, 1) {};
		\node [style=none] (52) at (5, 1) {};
		\node [style=none] (53) at (4.25, 1.25) {};
		\node [style=none] (54) at (7, 1.25) {};
		\node [style=none] (55) at (7, 1.75) {};
		\node [style=none] (56) at (7.5, 1.75) {};
		\node [style=none] (57) at (7.25, 1.25) {};
		\node [style=none] (58) at (7.5, 1.5) {};
		\node [style=none] (59) at (10, 1.5) {};
		\node [style=none] (60) at (10.5, 1.5) {};
		\node [style=none] (61) at (10, 1) {};
		\node [style=none] (62) at (10.5, 1) {};
		\node [style=none] (63) at (9.75, 1.25) {};
		\node [style=none] (64) at (9.75, 1.75) {};
		\node [style=none] (65) at (10.25, 1.75) {};
		\node [style=none] (66) at (10, 1.25) {};
		\node [style=none] (67) at (10.25, 1.5) {};
		\node [style=none] (68) at (-4.75, 1.5) {};
		\node [style=none] (69) at (-3.75, 1.5) {};
		\node [style=none] (70) at (-1, 1.5) {};
		\node [style=none] (71) at (0, 1.5) {};
		\node [style=none] (72) at (2.75, 1.5) {};
		\node [style=none] (73) at (3.75, 1.5) {};
		\node [style=none] (74) at (5.5, 1.5) {};
		\node [style=none] (75) at (6.5, 1.5) {};
		\node [style=none] (76) at (8.25, 1.5) {};
		\node [style=none] (77) at (9.25, 1.5) {};
		\node [style=none] (78) at (1.25, 3.75) {\tiny \textsf{3 x 3 x 4}};
		\node [style=none] (79) at (4.5, 3.75) {\tiny \textsf{2 x 2 x 1}};
		\node [style=none] (80) at (7.25, 3.75) {\tiny \textsf{1 x 1 x 2}};
		\node [style=none] (81) at (10, 3.75) {\tiny \textsf{1 x 1 x 2}};
		\node [style=none] (82) at (-4.25, 0) {\tiny \textsf{convolutional}};
		\node [style=none] (83) at (-0.5, 0) {\tiny \textsf{pooling}};
		\node [style=none] (86) at (-4.25, -0.5) {\tiny \textsf{layer}};
		\node [style=none] (87) at (3.25, 0) {\tiny \textsf{convolutional}};
		\node [style=none] (88) at (3.25, -0.5) {\tiny \textsf{layer}};
		\node [style=none] (89) at (6, 0) {\tiny \textsf{convolutional}};
		\node [style=none] (90) at (6, -0.5) {\tiny \textsf{layer}};
		\node [style=none] (91) at (8.75, 0) {\tiny \textsf{dense}};
		\node [style=none] (92) at (8.75, -0.5) {\tiny \textsf{layer}};
		\node [style=none] (93) at (-8.75, 0) {\tiny \textsf{input}};
		\node [style=none] (94) at (-8.75, -0.5) {\tiny \textsf{layer}};
		\node [style=none] (95) at (-8.25, 1.5) {};
		\node [style=none] (96) at (-9.25, 1.5) {};
		\node [style=none] (97) at (-6.5, 3.75) {\tiny \textsf{7 x 7 x 1}};
		\node [style=none] (98) at (-4.25, -1.25) {\tiny $\textsf{(3 x 3 x 1) x 4}$};
		\node [style=none] (99) at (-0.5, -1.25) {\tiny $(\textsf{spatial extent: 2})$};
		\node [style=none] (100) at (-0.5, -1.75) {\tiny $(\textsf{stride: 1})$};
		\node [style=none] (101) at (3.25, -1.25) {\tiny $\textsf{(2 x 2 x 4) x 1}$};
		\node [style=none] (102) at (6, -1.25) {\tiny $\textsf{(2 x 2 x 1) x 2}$};
		\node [style=none] (103) at (8.75, -1.25) {\tiny $\textsf{(2 x 2)}$};
		\node [style=none] (104) at (12, 1.5) {};
		\node [style=none] (105) at (11, 1.5) {};
		\node [style=none] (106) at (-0.5, -0.5) {\tiny \textsf{layer}};
	\end{pgfonlayer}
	\begin{pgfonlayer}{edgelayer}
		\draw (0.center) to (1.center);
		\draw (3.center) to (2.center);
		\draw (2.center) to (0.center);
		\draw (1.center) to (3.center);
		\draw (4.center) to (5.center);
		\draw (7.center) to (6.center);
		\draw (6.center) to (4.center);
		\draw (5.center) to (7.center);
		\draw (10.center) to (9.center);
		\draw (9.center) to (8.center);
		\draw (10.center) to (11.center);
		\draw (8.center) to (12.center);
		\draw (14.center) to (17.center);
		\draw (14.center) to (13.center);
		\draw (13.center) to (15.center);
		\draw (15.center) to (16.center);
		\draw (19.center) to (20.center);
		\draw (20.center) to (21.center);
		\draw (19.center) to (22.center);
		\draw (22.center) to (23.center);
		\draw (25.center) to (26.center);
		\draw (28.center) to (27.center);
		\draw (27.center) to (25.center);
		\draw (26.center) to (28.center);
		\draw (31.center) to (30.center);
		\draw (30.center) to (29.center);
		\draw (31.center) to (32.center);
		\draw (29.center) to (33.center);
		\draw (35.center) to (38.center);
		\draw (35.center) to (34.center);
		\draw (34.center) to (36.center);
		\draw (36.center) to (37.center);
		\draw (39.center) to (40.center);
		\draw (40.center) to (41.center);
		\draw (39.center) to (42.center);
		\draw (42.center) to (43.center);
		\draw (44.center) to (45.center);
		\draw (47.center) to (46.center);
		\draw (46.center) to (44.center);
		\draw (45.center) to (47.center);
		\draw (49.center) to (50.center);
		\draw (52.center) to (51.center);
		\draw (51.center) to (49.center);
		\draw (50.center) to (52.center);
		\draw (54.center) to (55.center);
		\draw (55.center) to (56.center);
		\draw (54.center) to (57.center);
		\draw (56.center) to (58.center);
		\draw (59.center) to (60.center);
		\draw (62.center) to (61.center);
		\draw (61.center) to (59.center);
		\draw (60.center) to (62.center);
		\draw (63.center) to (64.center);
		\draw (64.center) to (65.center);
		\draw (63.center) to (66.center);
		\draw (65.center) to (67.center);
		\draw [style=new edge style 0] (69.center) to (68.center);
		\draw [style=new edge style 0] (71.center) to (70.center);
		\draw [style=new edge style 0] (73.center) to (72.center);
		\draw [style=new edge style 0] (75.center) to (74.center);
		\draw [style=new edge style 0] (77.center) to (76.center);
		\draw [style=new edge style 0] (95.center) to (96.center);
		\draw [style=new edge style 0] (104.center) to (105.center);
	\end{pgfonlayer}
\end{tikzpicture}
  }
  \vspace{-0.5cm}
\end{figure}
\end{minipage}

The network above is constructed by \ascode{convNetwork}, whose type \lstinline{ConvNetwork} is the coproduct of five possible layer types found in a convolutional network:

\begin{lstlisting}
type ConvNetwork = (InputLayer :+: DenseLayer :+: ConvLayer :+: PoolLayer :+: ReLuLayer)

convNetwork :: Free ConvNetwork a
convNetwork = do
    denselayer (randMat2D 2 2)    |\,\,| (zeroMat1D 2)
    convlayer  (randMat4D 2 2 1 2) (zeroMat2D 2 1)
    convlayer  (randMat4D 2 2 4 1) (zeroMat2D 1 1)
    poollayer  2 1
    convlayer  (randMat4D 3 3 1 4) (zeroMat2D 1 1)
    inputlayer
\end{lstlisting}

\begin{figure}
  % \centering
  \begin{subfigure}{0.49\textwidth}
    \centering
    \includegraphics[width=1\columnwidth]{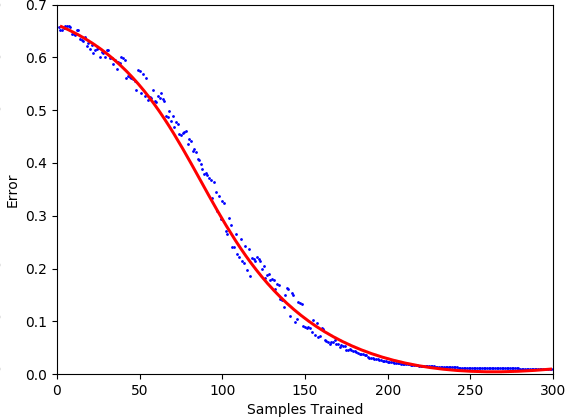}
    \vspace{-0.3cm}
  \end{subfigure}
  % \hspace*{0.1cm}
  \begin{subfigure}{0.49\textwidth}
    \centering
    \includegraphics[width=1\columnwidth]{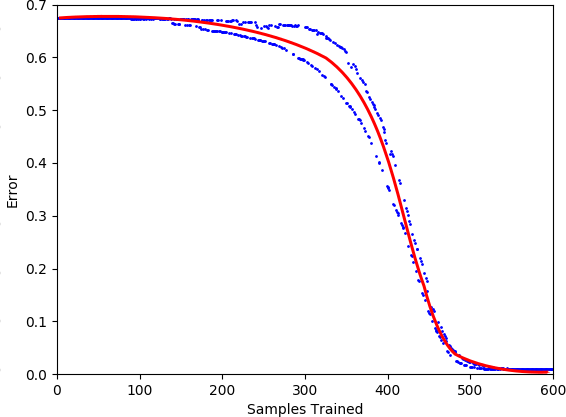}
    \vspace{-0.3cm}
    % \label{fig:fc_1400}
  \end{subfigure}
  \vspace{-0.1cm}
  \begin{subfigure}{0.49\textwidth}
    \centering
    \caption{Samples: 300, correlation coefficient: -0.747}
    \label{fig:conv_300}
  \end{subfigure}
  \begin{subfigure}{0.49\textwidth}
    \centering
    \caption{Samples: 600, correlation coefficient: -0.876}
    \label{fig:conv_600}
  \end{subfigure}
\caption{Training a convolutional neural network}
\label{fig:conv_results}
\vspace{-0.45cm}
\end{figure}

\noindent
In \figref{conv_results}, we see the change in error as the amount of trained samples increases, using sample sizes of 300 and 600.
The negative curvature demonstrates the convolutional neural network is successful in learning to classify the provided images. Two distinct streams of blue dots can also be observed in each graph, which represent the different paths of error induced by both of the sample types. The negative correlation coefficient is stronger in magnitude when using a sample size of 600, showing a better rate of convergence for the larger data set.

\subsection{Training a recurrent neural network}
\label{sec:lstm}

Recurrent networks are primarily used for classifying sequential time-series data in tasks such as text prediction, hand writing recognition or speech recognition. To demonstrate the functionality of our recurrent network implementation, we use DNA strands as data --- these can be represented using the four characters \textsf{a}, \textsf{t}, \textsf{c}, and \textsf{g}. Given a strand of five DNA characters, our network will attempt to learn the next character in the sequence.

\begin{figure}[H]
  \vspace{-0.1cm}
  \centering
  % \vspace{1cm}
\hspace{-1cm}
  \resizebox{0.6\textwidth}{!}{
    \centering\captionsetup[subfigure]{justification=centering}
    \begin{tikzpicture}
	\begin{pgfonlayer}{nodelayer}
		\node [style=new style 1] (0) at (-4, 2.5) {\tiny \textsf{c}};
		\node [style=new style 1] (1) at (-2, 2.5) {\tiny \textsf{c}};
		\node [style=new style 1] (2) at (0, 2.5) {\tiny \textsf{t}};
		\node [style=new style 1] (3) at (2, 2.5) {\tiny \textsf{a}};
		\node [style=new style 1] (4) at (4, 2.5) {\tiny \textsf{g}};
		\node [style=new style 2] (5) at (-4, 1) {\tiny \textsf{cell$_2^1$}};
		\node [style=new style 1] (15) at (-4, -2.25) {\tiny \textsf{a}};
		\node [style=new style 1] (16) at (-2, -2.25) {\tiny \textsf{c}};
		\node [style=new style 1] (17) at (0, -2.25) {\tiny \textsf{c}};
		\node [style=new style 1] (18) at (2, -2.25) {\tiny \textsf{t}};
		\node [style=new style 1] (19) at (4, -2.25) {\tiny \textsf{g}};
		\node [style=new style 2] (20) at (-2, 1) {\tiny \textsf{cell$_2^2$}};
		\node [style=new style 2] (21) at (0, 1) {\tiny \textsf{cell$_2^3$}};
		\node [style=new style 2] (22) at (2, 1) {\tiny \textsf{cell$_2^4$}};
		\node [style=new style 2] (23) at (4, 1) {\tiny \textsf{cell$_2^5$}};
		\node [style=new style 2] (24) at (-4, -0.75) {\tiny \textsf{cell$_1^1$}};
		\node [style=new style 2] (25) at (-2, -0.75) {\tiny \textsf{cell$_1^2$}};
		\node [style=new style 2] (26) at (0, -0.75) {\tiny \textsf{cell$_1^3$}};
		\node [style=new style 2] (27) at (2, -0.75) {\tiny \textsf{cell$_1^4$}}; 
		\node [style=new style 2] (28) at (4, -0.75) {\tiny \textsf{cell$_1^5$}};
	\end{pgfonlayer}
	\begin{pgfonlayer}{edgelayer}
		\draw [style=new edge style 0] (0) to (5);
		\draw [style=new edge style 0] (24) to (15);
		\draw [style=new edge style 0] (5) to (24);
		\draw [style=new edge style 0] (20) to (25);
		\draw [style=new edge style 0] (25) to (16);
		\draw [style=new edge style 0] (26) to (17);
		\draw [style=new edge style 0] (21) to (26);
		\draw [style=new edge style 0] (27) to (18);
		\draw [style=new edge style 0] (22) to (27);
		\draw [style=new edge style 0] (28) to (19);
		\draw [style=new edge style 0] (23) to (28);
		\draw [style=new edge style 0] (23) to (22);
		\draw [style=new edge style 0] (28) to (27);
		\draw [style=new edge style 0] (27) to (26);
		\draw [style=new edge style 0] (26) to (25);
		\draw [style=new edge style 0] (26) to (25);
		\draw [style=new edge style 0] (25) to (24);
		\draw [style=new edge style 0] (20) to (5);
		\draw [style=new edge style 0] (21) to (20);
		\draw [style=new edge style 0] (22) to (21);
		\draw [style=new edge style 0] (1) to (20);
		\draw [style=new edge style 0] (2) to (21);
		\draw [style=new edge style 0] (3) to (22);
		\draw [style=new edge style 0] (4) to (23);
	\end{pgfonlayer}
\end{tikzpicture}
  }
  \caption{Recurrent network}
  \label{fig:lstm_network_appendix}
  \vspace{-0.3cm}
\end{figure}
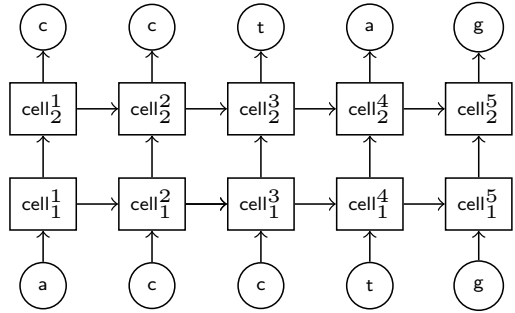

A diagram of the recurrent network used is shown in Figure~\ref{fig:lstm_network_appendix}, consisting of two layers of five cells. We omit its corresponding implementation, but note that training over this network structure is represented by a catamorphism over layers where the algebra is a catamorphism over the cells of the layer.

\begin{figure}
  % \centering
  \begin{subfigure}{0.49\textwidth}
    \centering
    \includegraphics[width=1\columnwidth]{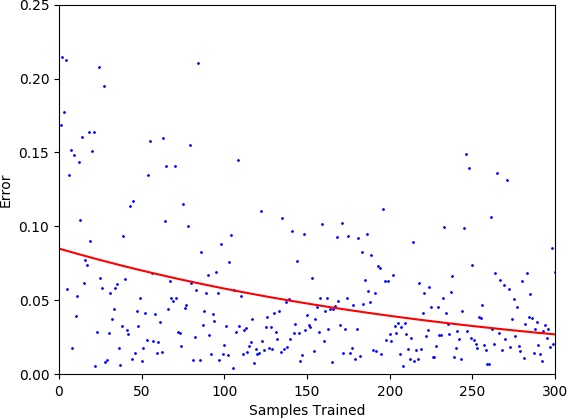}
    \vspace{-0.3cm}
  \end{subfigure}
  % \hspace*{0.1cm}
  \begin{subfigure}{0.49\textwidth}
    \centering
    \includegraphics[width=1\columnwidth]{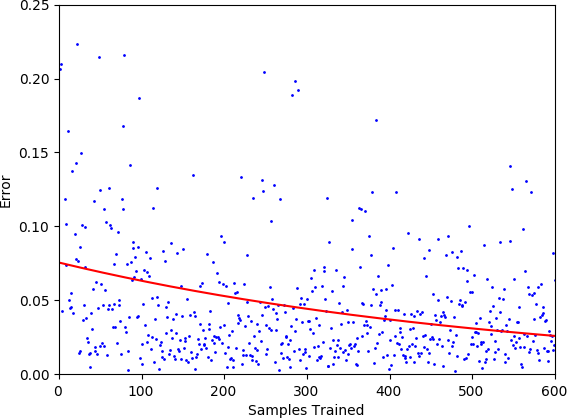}
    \vspace{-0.3cm}
    % \label{fig:fc_1400}
  \end{subfigure}
  \vspace{-0.1cm}
  \begin{subfigure}{0.49\textwidth}
    \centering
    \caption{Samples: 300, correlation coefficient: -0.314}
    \label{fig:lstm_300}
  \end{subfigure}
  \begin{subfigure}{0.49\textwidth}
    \centering
    \caption{Samples: 600, correlation coefficient: -0.254}
    \label{fig:lstm_600}
  \end{subfigure}
\caption{Training a recurrent neural network}
\label{fig:lstm-results}
\vspace{-0.45cm}
\end{figure}

In \figref{lstm-results}, we see the change in error as the amount of trained samples increases, using sample sizes of 300 and 600.
The negative curvature shown is less noticeable than the tests performed on previous networks, but still present. In contrast to the previous results, the correlation coefficient for the larger sample size of 600 is lower in magnitude than for the sample size of 300, perhaps showing convergence early on; achieving more conclusive results would require a more informed approach to the architecture and design of recurrent networks.

\end{document}